

\font\twelverm=cmr10 scaled 1200    \font\twelvei=cmmi10 scaled 1200
\font\twelvesy=cmsy10 scaled 1200

\skewchar\twelvei='177   \skewchar\twelvesy='60



             \font\tencsc=cmcsc10
             \font\fourteenrm=cmr10 scaled\magstep2
\font\seventeenrm=cmr10 scaled\magstep3
\font\twentyrm=cmr10 scaled 2073        


\def\tenpoint{
  \normalbaselineskip=12pt plus 0.1pt minus 0.1pt
  \parskip=6pt plus 4pt minus 4pt
  \abovedisplayskip 12pt plus 3pt minus 9pt
  \belowdisplayskip 12pt plus 3pt minus 9pt
  \abovedisplayshortskip 0pt plus 3pt
  \belowdisplayshortskip 7pt plus 3pt minus 4pt
  \smallskipamount=3pt plus1pt minus1pt
  \medskipamount=6pt plus2pt minus2pt
  \bigskipamount=12pt plus4pt minus4pt
  \textfont0=\tenrm   \scriptfont0=\sevenrm   \scriptscriptfont0=\fiverm
  \textfont1=\teni    \scriptfont1=\seveni    \scriptscriptfont1=\fivei
  \textfont2=\tensy   \scriptfont2=\sevensy   \scriptscriptfont2=\fivesy
  \textfont3=\tenex   \scriptfont3=\tenex     \scriptscriptfont3=\tenex
  \textfont\itfam=\tenit
  \textfont\slfam=\tensl
  \textfont\bffam=\tenbf \scriptfont\bffam=\sevenbf
  \scriptscriptfont\bffam=\fivebf
  \def\rm{\fam0\tenrm}          \def\byg{\fam0\twelverm}
  \def\vbig{\fam0\fourteenrm}   \def\vvbig{\fam0\seventeenrm}
  \def\vvvbig{\fam0\twentyrm}   \def\it{\fam\itfam\tenit}
  \def\sl{\fam\slfam\tensl}     \def\bf{\fam\bffam\tenbf}
  \def\mit{\fam 1}              \def\cal{\fam 2}
  \def\tt{\tentt}               \def\csc{\tencsc}
  \normalbaselines\rm
}


\def\junk{\parskip=0pt \obeylines\def\\{\par}}
\def\beginlinemode{\endmode\begingroup
        \junk\def\endmode{\par\endgroup}}
\def\beginparmode{\endmode\begingroup
        \def\endmode{\par\endgroup}}

\def\body{\beginparmode}

\let\endmode=\par

\def\endpage{\vfill\eject}

{\obeylines\gdef\
{ }}

\def\singlespace{\baselineskip=\normalbaselineskip}
\def\oneandathirdspace{\baselineskip=\normalbaselineskip
        \multiply\baselineskip by 4 \divide\baselineskip by 3}
\def\oneandahalfspace{\baselineskip=\normalbaselineskip
        \multiply\baselineskip by 3 \divide\baselineskip by 2}
\def\doublespace{\baselineskip=\normalbaselineskip
        \multiply\baselineskip by 2}

\newcount\firstpageno
\firstpageno=2
\footline={\ifnum\pageno<\firstpageno{\hfil}\else{\hfil\folio\hfil}\fi}

\let\rawfootnote=\footnote
\def\footnote#1#2{{\rm\singlespace\parindent=0pt\parskip=0pt
  \rawfootnote{#1}{#2\hfill\vrule height 0pt depth 6pt width 0pt}}}
\def\raggedcenter{
  \leftskip=4em plus 12em \rightskip=\leftskip
  \parindent=0pt \parfillskip=0pt \spaceskip=.3333em \xspaceskip=.5em
  \pretolerance=9999 \tolerance=9999
  \hyphenpenalty=9999 \exhyphenpenalty=9999}


\def\today{
  \ifcase\month\or January\or February\or March\or
  April\or May\or June\or July\or August\or
  September\or October\or November\or December\fi
  \space\number\day, \number\year}


\def\ppt#1{\rightline{\rm UdeM-LPN-TH-#1}}

\def\title{\null\vskip 3pt plus 0.2fill
        \beginlinemode \doublespace \raggedcenter \bf\byg}

\def\author{                    
  \vskip 8pt plus 0.2fill \beginlinemode
  \singlespace \raggedcenter\rm}

\def\affil{                     
  \vskip 3pt plus 0.1fill \beginlinemode
  \oneandathirdspace \raggedcenter \sl}

\def\lpn{Laboratoire de Physique Nucl\'eaire\\
	Universit\'e de Montr\'eal\\Montr\'eal, Qu\'e H3C 3J7\\
	Canada}

\def\abstract{                   
  \vskip 3pt plus 0.3fill \beginparmode
  \singlespace}

\def\endtopmatter{               
  \endpage\body}


\magnification\magstep1
\hsize=6.5truein
\vsize=8.9truein
\clubpenalty=500                
\widowpenalty=500               
\def\\{\cr}
\tenpoint
\oneandahalfspace
\overfullrule=0pt       

\def\head#1{                    
  \goodbreak\bigskip\bigskip    
  {\immediate\write16{#1}
   \raggedcenter \uppercase{#1}\par}
   \nobreak\bigskip\nobreak}

\def\beginitems{
\par\medskip\bgroup\def\i##1 {\item{##1}}\def\ii##1 {\itemitem{##1}}
\leftskip=36pt\parskip=0pt}
\def\enditems{\par\egroup}




\def\references                 
  {\head{References}            
   \beginparmode\singlespace
   \frenchspacing \parindent=0pt \leftskip=1truecm
   \parskip=4pt plus 3pt \everypar{\hangindent=\parindent}}

\def\refto#1{$^{#1}$}           
\gdef\refis#1{\item{#1.\ }}                     

\gdef\journal#1, #2, #3, 1#4#5#6{               
    {\sl #1~}{\bf #2}, #3 (1#4#5#6)}            

\def\pr{\journal Phys. Rev., }

\def\prb{\journal Phys. Rev. B, }

\def\prd{\journal Phys. Rev. D, }
\def\prl{\journal Phys. Rev. Lett., }

\def\np{\journal Nucl. Phys., }
\def\pl{\journal Phys. Lett., }

\def\annp{\journal Ann. Phys. (N.Y.), }
\def\mpl{\journal Mod. Phys. Lett., }
\def\ijmp{\journal Int. Jour. Mod. Phys., }

\def\ref#1{Ref.~#1}                     
\def\Ref#1{Ref.~#1}                     
\def\[#1]{[\cite{#1}]}
\def\cite#1{{#1}}
\def\(#1){(\call{#1})}
\def\call#1{{#1}}
\def\endreferences{\body}


\def\figurecaptions             
  {\endpage
   \head{Figure Captions}
   \beginparmode\singlespace
   \parindent=0pt \leftskip=1truecm
   \parskip=4pt plus 3pt \everypar={\hangindent=\parindent}}

\def\endfigurecaptions{\body}


\def\bye  {\endmode\vfill\supereject\end}
\singlespace
%
%
\def\cf{{\it cf.\ }}
\def\eg{{\it e.g.},\ }

\def\ie{{\it i.e.},\ }

\def\etalc{{\it et al.},\ }

%
%
\def\a{\alpha}
\def\be{{\beta}}
\def\ga{{\gamma}}

\def\eps{\epsilon}

\def\la{\lambda}

\def\om{{\omega}}

\def\La{{\Lambda}}

%
%

\def\DD{{\cal D}}

\def\LL{{\cal L}}

\def\Tr{{\rm Tr}}

\def\3he{{$^3{\rm He}$}}

\def\bfk{{\bf k}}

\def\bfq{{\bf q}}

\def\bfR{{\bf R}}

\def\bfx{{\bf x}}

\def\bfy{{\bf y}}

%
%
\def\sla{\raise.16ex\hbox{$/$}\kern-.57em}
\def\Dsl{\kern.2em\raise.16ex\hbox{$/$}\kern-.77em\hbox{$D$}}
\def\parsl{\sla\hbox{$\partial$}}

\def\Asl{\kern.2em\raise.16ex\hbox{$/$}\kern-.77em\hbox{$A$}}
\def\psl{\sla\hbox{$p$}}
\def\ksl{\sla\hbox{$k$}}
\def\qsl{\sla\hbox{$q$}}
\def\gtwid{\mathrel{\raise.3ex\hbox{$>$\kern-.75em\lower1ex\hbox{$\sim$}}}}
\def\ltwid{\mathrel{\raise.3ex\hbox{$<$\kern-.75em\lower1ex\hbox{$\sim$}}}}

%
%

\def\you#1{{\rm U}(#1)}

%
%

\def\aboverel#1\above#2{\mathrel{\mathop{#2}\limits^{#1}}}
\def\beneathrel#1\under#2{\mathrel{\mathop{#2}\limits_{#1}}}
\def\frac#1#2{{#1 \over #2}}
\def\half{{\frac 12}}

\def\12{{1\over2}}

\def\part{\partial}

\def\square{\kern1pt\vbox{\hrule height 1.2pt\hbox{\vrule width 1.2pt\hskip 3pt
   \vbox{\vskip 6pt}\hskip 3pt\vrule width 0.6pt}\hrule height 0.6pt}\kern1pt}

\def\low#1{\lower.5ex\hbox{${}_#1$}}
\def\tdot#1{\mathord{\mathop{#1}\limits^{\kern2pt\ldots}}}
%
%

\def\refto#1{$^{#1}$}

\def\leaderfill{\leaders\hbox to 1em{\hss.\hss}\hfill}

\catcode`@=11
\newcount\r@fcount \r@fcount=0
\newcount\r@fcurr
\immediate\newwrite\reffile
\newif\ifr@ffile\r@ffilefalse
\def\w@rnwrite#1{\ifr@ffile\immediate\write\reffile{#1}\fi\message{#1}}

\def\writer@f#1>>{}
\def\referencefile{
  \r@ffiletrue\immediate\openout\reffile=\jobname.ref%
  \def\writer@f##1>>{\ifr@ffile\immediate\write\reffile%
    {\noexpand\refis{##1} = \csname r@fnum##1\endcsname = %
     \expandafter\expandafter\expandafter\strip@t\expandafter%
     \meaning\csname r@ftext\csname r@fnum##1\endcsname\endcsname}\fi}%
  \def\strip@t##1>>{}}

\def\citeall#1{\xdef#1##1{#1{\noexpand\cite{##1}}}}
\def\cite#1{\each@rg\citer@nge{#1}}     

\def\each@rg#1#2{{\let\thecsname=#1\expandafter\first@rg#2,\end,}}
\def\first@rg#1,{\thecsname{#1}\apply@rg}       
\def\apply@rg#1,{\ifx\end#1\let\next=\relax
\else,\thecsname{#1}\let\next=\apply@rg\fi\next}

\def\citer@nge#1{\citedor@nge#1-\end-}  
\def\citer@ngeat#1\end-{#1}
\def\citedor@nge#1-#2-{\ifx\end#2\r@featspace#1 
  \else\citel@@p{#1}{#2}\citer@ngeat\fi}        
\def\citel@@p#1#2{\ifnum#1>#2{\errmessage{Reference range #1-#2\space is bad.}%
    \errhelp{If you cite a series of references by the notation M-N, then M and
    N must be integers, and N must be greater than or equal to M.}}\else%
 {\count0=#1\count1=#2\advance\count1
by1\relax\expandafter\r@fcite\the\count0,%
  \loop\advance\count0 by1\relax
    \ifnum\count0<\count1,\expandafter\r@fcite\the\count0,%
  \repeat}\fi}

\def\r@featspace#1#2 {\r@fcite#1#2,}    
\def\r@fcite#1,{\ifuncit@d{#1}
    \newr@f{#1}%
    \expandafter\gdef\csname r@ftext\number\r@fcount\endcsname%
                     {\message{Reference #1 to be supplied.}%
                      \writer@f#1>>#1 to be supplied.\par}%
 \fi%
 \csname r@fnum#1\endcsname}
\def\ifuncit@d#1{\expandafter\ifx\csname r@fnum#1\endcsname\relax}%
\def\newr@f#1{\global\advance\r@fcount by1%
    \expandafter\xdef\csname r@fnum#1\endcsname{\number\r@fcount}}

\let\r@fis=\refis                       
\def\refis#1#2#3\par{\ifuncit@d{#1}
   \newr@f{#1}%
   \w@rnwrite{Reference #1=\number\r@fcount\space is not cited up to now.}\fi%
  \expandafter\gdef\csname r@ftext\csname r@fnum#1\endcsname\endcsname%
  {\writer@f#1>>#2#3\par}}

\def\ignoreuncited{
   \def\refis##1##2##3\par{\ifuncit@d{##1}%
     \else\expandafter\gdef\csname r@ftext\csname
r@fnum##1\endcsname\endcsname%
     {\writer@f##1>>##2##3\par}\fi}}

\def\r@ferr{\endreferences\errmessage{I was expecting to see
\noexpand\endreferences before now;  I have inserted it here.}}
\let\r@ferences=\references
\def\references{\r@ferences\def\endmode{\r@ferr\par\endgroup}}

\let\endr@ferences=\endreferences
\def\endreferences{\r@fcurr=0
  {\loop\ifnum\r@fcurr<\r@fcount
    \advance\r@fcurr by 1\relax\expandafter\r@fis\expandafter{\number\r@fcurr}%
    \csname r@ftext\number\r@fcurr\endcsname%
  \repeat}\gdef\r@ferr{}\endr@ferences}


\let\r@fend=\endpaper\gdef\endpaper{\ifr@ffile
\immediate\write16{Cross References written on []\jobname.REF.}\fi\r@fend}

\catcode`@=12

\citeall\refto          
\citeall\ref            %
\citeall\Ref            %

\def\leff{\LL_{\rm eff}}
\def\seff{S_{\rm eff}}
\def\veff{V_{\rm eff}}
\def\skin{S^{\rm kin}}
\def\vt{v_{\rm T}}
\def\tkt{T_{\rm KT}}
\def\stuff{\left(1-{\beta \vt\over \sinh(\beta \vt)}\right)}
\ignoreuncited
\ppt{147}
\title {Superconductivity in $2+1$ dimensions via Kosterlitz-Thouless
Mechanism: Large-N and Finite-Temperature Analyses}
\author R. MacKenzie, P.K. Panigrahi\footnote{$^*$}{Address after 1 Oct,
1993: Department of Physics, University of Hyderabad, Hyderabad, India,
500134.} and S. Sakhi
\affil \lpn
\abstract
\centerline{Abstract}
We analyse a $2+1$ dimensional model with charged,
relativistic fermions interacting through a
four-Fermi term. Taking advantage of its large-$N$ renormalizability,
the various phases of this model are studied at finite
temperature and beyond the leading order in $1/N$. Although the vacuum
expectation value (VEV) of a charged order parameter is zero at any non-zero
temperature, the model nevertheless exhibits a rich phase structure
in the strong coupling r\'egime, because
of the non-vanishing VEV of a neutral order parameter and due to the
non-trivial dynamics of the vortex excitations on the plane. These
are: a confined-vortex phase which is superconducting at low
temperatures, an intermediate-temperature phase with deconfined vortices,
and a high-temperature phase, where the neutral order parameter vanishes.
The manifestation of superconductivity at low-temperatures and its
disappearance above a critical temperature is explicitly
shown to be due to the vortex confinement/deconfinement mechanism of
Kosterlitz and Thouless. The ground state does not break
parity or time-reversal
symmetries and the ratio of the energy gap to $T_c$
is bigger than the conventional BCS value, for $N\ltwid 22$.
\endtopmatter

\head{I. INTRODUCTION}

Of late, there has been considerable interest in planar field theoretical
models due to their relevance to
condensed matter systems.\refto{fradkin,balercmorsri}
Furthermore, the
recent observation\refto{roswarpar} of the renormalizability of four-Fermi
couplings in
$2+1$-dimensions in large-$N$
perturbation theory,\refto{thooft,coleman4} has given a tremendous boost to the
systematic
analysis of the phase
structures of various interacting theories.\refto{roswarpar,gatkovros,yamich}
The possibility of
observing a superconducting ground state in some of these phases is obviously
of great practical and theoretical interest.
This is due to the fact that the conventional explanation of this
phenomenon based on the spontaneous breaking of a $U(1)$
symmetry may not be applicable to a lower-dimensional world.

For quite some
time, it has been known from the works
of Hohenberg,\refto{hohenberg} Mermin and Wagner\refto{merwag} and
Coleman\refto{coleman3}
that breaking of a continuous symmetry
(SSB) does not take place in $1+1$ dimensions
due to infrared divergences. This is strictly true in the absence of
long-range interactions. Similar infrared singularities also arise in $2+1$
dimensions at finite temperature, making the vacuum expectation value (VEV)
of
a charged order parameter vanish for non-zero temperatures. It is not very
dif\-fi\-cult
to identify the source of this problem. The low-energy field theory
in the broken symmetry phase is governed by a massless Bose field $\phi$
(the Goldstone mode), which in the imaginary-time formalism\refto{matsubara}
can be expanded
as $\phi{(\bf x,\tau)}=\sum_n\phi_n({\bf x})e^{i\omega_n\tau}$, where the
energy  takes on values
$\omega_n={2n\pi}/\beta$, $\beta$ being the inverse temperature. The $n=0$
mode is governed by ${S[\phi_0]=\int_0^\beta d\tau
\int d^2x\LL{(\phi_0,\partial \phi_0)}= {\beta \int d^2x
\LL{(\phi_0,\partial \phi_0)}}}$, a purely two-dimensional action.
Hence, it is not surprising that the afflictions of $1+1$ dimensional
theories
will manifest themselves
here. However, it is worth mentioning that a neutral order
parameter (possibly breaking a discrete symmetry) can have a VEV in $2+1$
dimensions at finite temperature.\refto{roswarpar}

In light of the above constraint, to achieve planar superconductivity,
one usually invokes a small interplanar coupling or takes recourse in
unconventional mechanisms, \eg anyon superconductivity.\refto
{laughlin1,fethanlau,chewilwithal,hoscha,wenzee2,banlyk,panraysak,ransalstr}
In this paper, an alternate model\refto{macpansak}
exhibiting superconductivity in the strong
coupling phase due to the vortex confinement mechanism of Kosterlitz and
Thouless (KT),\refto{kostho,berezinski} is analysed in some detail
via large-$N$ perturbation theory. The model under consideration is
a $2+1$ dimensional relativistic theory with a four-Fermi interaction
(reminiscent of the BCS theory\refto{barcoosch}) in
the presence of a weak and external
electromagnetic field. In the previous work\refto{macpansak}, the
low-energy effective action up to two-derivatives
has been explicitly computed
to leading order in $1/N$ using two independent
methods.\refto{macwilzee,aitfra}

The primary goal of the
present analysis is to go beyond the leading order in $1/N$ in order to
establish the renormalizability of the model and to show the absence of any
ultraviolet or
infrared divergences that could destabilize the vacuum structure.
The model exhibits three different phases as a
function of temperature and chemical potential. The low-temperature phase
exhibits perfect diamagnetism with a pole in the current-current correlator
because the vortices appear in tightly bound pairs by virtue of a logarithmic
attractive potential. It is worth mentioning that on a plane the
phase degree of freedom  of a complex order parameter is multivalued, thereby
giving rise to vortex configurations. At a temperature $\tkt $, the vortices
are deconfined; it is explicitly shown that this is also the temperature
where superconductivity vanishes. This result was anticipated, but not
proven, in some recent
works in the literature.\refto{kovros,dormav,leefis}
At a still higher temperature, the
neutral order parameter vanishes through a second order phase transition.

The motivation for analysing the above model lies in the appearance of
relativistic fermions as the relevant low-energy degrees of freedom in the
strong coupling limit of the Hubbard
and related models,\refto{maraff,shankar}
widely believed to be of relevance to the high-$T_c$  superconductors.
Furthermore, four-Fermi couplings also arise naturally  in the
above-mentioned models in the presence of doping.\refto{shankar,semwij1}
In the
absence of precise knowledge regarding the form and strength of these
couplings, we  consider a four-Fermi term of the BCS type, preserving the
relativistic invariance of the original Lagrangian. The method of analysis
can be easily generalised to other types of couplings.
Finally, although not conclusive,  there is
some experimental indication as to the KT nature of the phase  transition
in the high-$T_c$ materials,\refto{staforaya}
giving  urgency to the analyses of
various models exhibiting such behaviour.

The goal of the
study is to capture the generic properties of this class of models  using
the large-$N$ renormalizability of the  four-Fermi couplings in  $2+1$
dimensions and find the similarities and  dissimilarities
with the standard BCS theory. Large-$N$ perturbation theory is inherently
non-perturbative: a given order in this approximation scheme
represents a class of diagrams with various powers of the conventional
weak-coupling expansion parameter. The different orders in this perturbation
theory
also have nice geometrical interpretation in terms of Riemann surfaces;
${\it e.g,}$
the leading order expression is the sum of all planar diagrams (genus zero
surface).\refto{coleman4} In the present context the large-$N$ expansion
is ideal because of the fact that the weak-coupling perturbation is not
renormalizable as can be easily seen from power counting.
Interestingly, our analysis reveals that unlike
the BCS case, superconductivity
in the model under consideration occurs in the strong coupling phase.

The organization of the paper is as follows. In Sec. II, we give a brief
review of the derivation of the Landau-Ginzburg effective action
in the BCS theory and proceed to do the same in our model.
We start with the analysis of
the effective potential $\veff$ and show that to leading order in
the $1/N$ expansion a neutral order parameter
can have a non-vanishing VEV up to a critical temperature $T_0$.
Subsequently,
the low-energy effective action is computed and used to show that
the VEV of
a charged order parameter at any non-zero temperature vanishes, because
of the presence of a massless mode. In Sec. III, we discuss the
renormalizability of the model after taking into account the
next-to-leading order corrections.
We also show that the VEV of the neutral order parameter is
not destabilized by quantum fluctuations.
Sec. IV is devoted to the study of the
occurrence of superconductivity
in this model, taking into account the non-trivial
dynamics of the phase degrees of freedom. Finally, we conclude in Sec.
V with
some comments and future directions of work.

\vfill\eject
\head{II. ONE-LOOP EFFECTIVE ACTION}

In the discussions of BCS superconductivity by field
theorists,\refto{sakita,schakel}
a common
approach is the following. One starts from
the second-quantized BCS Lagrangian, including an effective
electron-electron interaction due to phonon exchange:
$$
\LL_{\rm BCS}
=\psi^\dagger_\uparrow\left(i\partial_t-\eps(p)\right)\psi_\uparrow
+\psi^\dagger_\downarrow\left(i\partial_t-\eps(p)\right)\psi_\downarrow
-\la\psi^\dagger_\uparrow\psi^\dagger_\downarrow
\psi_\downarrow\psi_\uparrow.
\eqno(1)
$$
Here $\eps(p)$ is the kinetic energy, including a chemical potential and
coupling to electromagnetism, if desired.
This has the obvious $\you1$ phase symmetry $\psi\to e^{i\theta}\psi$.
One then argues that the low-energy behaviour of the theory is best
described in terms of a condensate field rather than the fermions.
Physically this is connected to the fact that there is a gap in the fermion
spectrum below the critical temperature, as a result of which
long-wavelength excitations
of the condensate are much cheaper to create than fermionic excitations.

In order to achieve an effective theory for the condensate, one performs a
Hubbard-Stratonovich\refto{hubbard1,stratonovich} transformation, yielding
$$
\LL_1=\psi^\dagger_\uparrow(i\partial_t-\eps(p))\psi_\uparrow
+\psi^\dagger_\downarrow(i\partial_t-\eps(p))\psi_\downarrow
+\sqrt{\la g}(\phi^*\psi_\downarrow\psi_\uparrow
+\phi\psi^\dagger_\uparrow\psi^\dagger_\downarrow)+g\phi^*\phi.
\eqno(2)
$$

The advantage of (2) is that its
bilinear nature in the fermion fields
permits one to eliminate the fermion, at
least formally. In the language of path integrals, the generating functional
of (2) is
$$
Z=\int\DD\psi^\dagger\DD\psi\DD\phi^*\DD\phi\,e^{i S_1},
\eqno(3)
$$
where $S_1=\int d^4x\,\LL_1$.
Notice that the functional integral over the scalar field can be done,
yielding an ``effective action'' for the fermion which is none other than
the integral of (1). Alternatively,
one can perform the fermion functional integral:
$$
Z=\int\DD\phi^*\DD\phi\,e^{i \seff(\phi^*,\phi)},
$$
where
$$
e^{i \seff[\phi^*,\phi]}=\int\DD\psi^\dagger\DD\psi e^{i S_1}
\eqno(4a)
$$
and
$$
\seff=-i\Tr\, \log\pmatrix{p_0-\eps(p)&-\sqrt{\la g}\phi\cr
				\sqrt{\la g}\phi^*&p_0+\eps(p)\cr}
+\int d^4x\,g\phi^*\phi.
\eqno(4b)
$$
Here the matrix, if sandwiched between
$\psi^\dagger=(\psi^\dagger_\uparrow,\psi_\downarrow)$ and
$\psi=(\psi_\uparrow,\psi^\dagger_\downarrow)^T$, gives the
fermion-dependent
part of (2).

Next, the effective action can be evaluated in a gradient expansion. The
lowest-order term, with no derivatives, is the effective potential for
the condensate; at finite temperature, the result is
$$
\veff\sim a\log{T\over T_c}\phi^*\phi +b(\phi^*\phi)^2,
\eqno(5)
$$
where $a,\ b$ are positive parameters, the details of which are not
important to us. One sees from the effective potential
that $\phi$ attains a nonzero VEV below a
certain critical temperature. When that happens, electromagnetism is broken
and superconductivity results. The phase transition is of second order.

In this paper, we pursue a similar analysis in a
relativistic setting taking in to account the peculiarities of $2+1$
dimensional space-time.

Probably the first situation
in which a calculation similar to the above was
done in a relativistic setting was the work of Nambu and
Jona-Lasinio.\refto{namjon} This work, which is actually approximately
concurrent with BCS, introduced the idea of dynamical symmetry breaking in
the context of the pion-nucleon system. A four-Fermi interaction was shown
to break chiral symmetry, leading to what are now known as Nambu-Goldstone
bosons, which Nambu and Jona-Lasinio identified with the pions.

Another
more recent field theory which mimics the BCS theory as discussed above
is the Gross-Neveu model,\refto{gronev}
a 1+1-dimensional model described by the
following Lagrangian:
$$
\LL_{\rm GN}=\bar\psi_i i\parsl\psi_i
+{g^2\over 2}\left (\bar\psi_i\psi_i\right)^2.
\eqno(6)
$$
Again one often introduces an auxiliary field to remove the four-Fermi
interaction, replacing the last term in (6) with
$$
g\phi\bar\psi_i\psi_i-\half\phi^2.
\eqno(7)
$$
It is well known that the
discrete chiral transformation $\psi\to\ga_5\psi$ is broken spontaneously.
This symmetry breaking, together with the fact that the model is
asymptotically free, makes it a good testing ground for QCD (although the
fact
that it is in 1+1 dimensions makes the connection rather tenuous).

One obvious difference between the Gross-Neveu model and the BCS theory is
that the ``condensate'' $\phi$ in (7) is a real (uncharged) scalar, while
that in BCS is doubly charged. One can
of course couple a charged scalar to fermions via an interaction of the
form
$\bar\psi\psi|\phi|^2,$
but this is not terribly reminiscent of BCS.
In order to find a closer analog we must find a doubly-charged bilinear in
the fermion. This can be done using
the charge-conjugate fermion field,
$\psi^c=C\bar\psi^T$, C being the charge conjugation matrix.

The appropriate relativistic
generalization of (1), which we will analyze in detail, is thus
$$
\LL=\bar\psi_\alpha (i\parsl-e\Asl)\psi_\alpha-{1\over 4\lambda N}
\bar\psi_\alpha\psi_\alpha^c\bar\psi_\beta^c\psi_\beta.
\eqno(8)
$$
Here, $\alpha, \beta$ range from 1 to $N$, $N$ being the parameter of the
large-$N$ expansion, and repeated indices imply summation.
Furthermore, the coupling constant $1/\lambda$ has
canonical dimension (mass)$^{-1}$ and hence conventional perturbation
theory is non-renormalizable. However, as mentioned earlier this type of
four-Fermi coupling is
renormalizable in large-$N$ perturbation theory,\refto{roswarpar}
making it an
ideal tool to analyse the phase structure.
Coupling to the photon field endows the model with a $\you1$ gauge
symmetry.
Like the BCS theory, it is assumed that the short range attractive force
represented by the four-Fermi term,
originating from the underlying vibrations
of the microscopic theory, dominates the Coulomb repulsion between the
electrons at low-temperatures so that the net interaction is attractive.

In what follows, we will remove the four-Fermi interaction in favour of a
Yukawa-type coupling to an auxiliary charged scalar field $\phi$ and then
perform a derivative expansion of the effective theory for $\phi$.
As a first step, the effective potential will be computed to leading order
in $1/N$ to study the question of radiative breaking of the $\you1$
symmetry {\it \`a la} Coleman-Weinberg.\refto{colwei} After
answering this
question in the affirmative, finite temperature effects will then be
incorporated to show that the VEV of $\phi$ is zero
when
$T \neq 0$, although the neutral order parameter $|\phi|$ has a
non-zero VEV.
At a still higher temperature the neutral order parameter vanishes,
the phase transition being of second order.

Were $\phi$ a true dynamical field, rather than an auxiliary
field, we would argue that the two-derivative term induced by the fermion
loop serves only to renormalize the already-present kinetic term for
$\phi$. However, being an auxiliary
field, it has no kinetic term and the induced
two-derivative term is physically significant as it stands. In order for
the expansion up to two derivatives to be sensible, therefore, the
coefficient of this term must be positive. We check this, and find that it
is.

We use a diagrammatic approach to compute the derivative
expansion,\refto{macwilzee} making use of the interpretation of the
effective action in terms of one-particle-irreducible Green's functions.
The approach is essentially a generalization of the method devised by
Coleman and Weinberg\refto{colwei}
to compute the effective potential. An alternative
approach using an expansion of the trace-log representation
of the effective action\refto{aitfra} is employed in Sec.
III as a check, and also to compute
the next-to-leading order effects.

It is worth pointing out that the Lagrangian under consideration
here, with additional scalar self-couplings, has been analysed before
for the purpose of finding zero-modes of the Dirac equation in the presence
of a vortex, either without\refto{jacros,weinberg} or
with\refto{ganhat,macmat} a
Chern-Simons term. Also, radiative symmetry breaking in various similar models
in 2+1 dimensions has been discussed in Ref.~\cite{babpanram}.
Hence our investigation supplements these previous
works.

Introducing auxiliary fields $\phi$ and $\phi^*$ to decouple the four-Fermi
term in (8), we get:
$$
\LL=\bar\psi_\alpha (i\parsl-e\Asl)\psi_\alpha+{1\over2}
\left(\phi^*\bar\psi^c_\alpha\psi_\alpha
-\phi\bar\psi_\alpha\psi^c_\alpha\right)-\lambda N\phi^*\phi.
\eqno(9)
$$
To maintain gauge invariance of the Yukawa term
in (9), the charge of the complex field is $2e$.

The effective action for the scalar and gauge fields
is attained by integrating over
the fermions:
$$
e^{i \seff[A,\phi,\phi^*]}=\int\DD\bar\psi\DD\psi\,e^{i S}.
$$
In fact, the dependence of $\seff$ on $A$ can be deduced from gauge
invariance, so often in what follows we set $A$ to zero.

A nonlocal object, $\seff$ cannot be computed exactly. However, if our
goal is to describe the low-energy, long-wavelength excitations of the
condensate, it is reasonable to attempt a gradient expansion of
$\seff$. Indeed, the equivalent of the trace-log in (4) can be computed as
a derivative expansion directly.\refto{aitfra,babdaspan}
However, an approach which is perhaps more physical
is to appeal to the interpretation of the effective action as an object
which incorporates the effects of scalar field interactions mediated by
internal fermion lines. Thus, $\seff$ is the
generating functional of one-particle irreducible vertex functions:
$$
\eqalignno{
	\seff[\phi,\phi^*]
	&=-i\sum_n{1\over n!^2}
	\int{d^3p_1\over(2\pi)^3}{d^3q_1\over(2\pi)^3}\cdots
	{d^3p_{n}\over(2\pi)^3}{d^3q_n\over(2\pi)^3}
	(2\pi)^3\delta^3\left(\sum(p_i+q_i)\right)\cr
	&\quad\times\Gamma^{(n,n)}(p_1,\cdots,p_{n};q_1,\cdots,q_{n})
	\tilde\phi(p_1)\tilde\phi^*(q_1)\cdots
	\tilde\phi(p_{n})\tilde\phi^*(q_{n}).&(10)\cr}
$$
Alternatively, simply by recognizing
$\seff$ as a functional of $\phi$, it
can be expressed in terms of the gradients of
$\phi$ and $\phi^*$:
$$
\seff[\phi,\phi^*]=\int d^3x\,\left(-\veff(\phi)
+Z_1(\phi)\partial_\mu\phi\partial^\mu\phi^*
+Z_2(\phi)(\phi^*\partial_\mu\phi +\phi\partial_\mu\phi^*)^2
+\cdots\right).
\eqno(11)
$$
The key feature of the first expansion is that each term is calculable:
$\Gamma^{(n,n)}$ is the one-fermion-loop diagrams with $n$ external $\phi$'s
and $\phi^*$'s. On
the other hand, the second, a derivative expansion,
is a convenient way of capturing the
long-wavelength behaviour of the scalar field as induced by the fermion.
In what follows we will compute the first two terms in the
derivative expansion through judicious
combinations of the 1PI Green's functions.
To leading order the massless fermion propagator is
${i/\psl}$; the two vertices coming from $\phi$, $\phi^*$
couplings to fermion fields $\psi_i$
and $\psi^*_j$ are ${i}C^{-1}\delta_{ij}/2$ and
${i}C\delta_{ij}/2$ respectively.

We concentrate first on the effective potential.
By rewriting (11) in momentum space, $\veff$ can be calculated from
the Green's functions at zero
momentum since the other terms vanish in this limit. It is not difficult
to
see that
$$
\veff(\phi)=V^0+i\sum_{n=1}^\infty{(\phi^*\phi)^n\over n!^2}
\Gamma^{(n,n)}(0,\cdots0;0,\cdots0).
\eqno(12)
$$
Here, $V^0$ is the classical potential, \ie the last term in (9).
The sum over all zero-momentum diagrams appears formidable, but as shown
by
Coleman and Weinberg, it is actually quite doable. The only unusual twist
in the present case is actually in the Feynman rules: the
$\phi\bar\psi\psi^c$ coupling, for example, gives a vertex with an incoming
$\phi$ and two outgoing fermions. Formally, the effective potential is given by
the classical potential plus the sum
of diagrams depicted in Fig.~1, with external fields at zero momentum.

The above sum can be carried out, yielding
$$
\veff=V^{(0)}
+iN\int{d^3k\over(2\pi)^3}\log\left(1-{\phi^*\phi\over k^2}\right).
\eqno(14)
$$
$\veff$ is linearly divergent; we thus impose a cutoff $\La$.
Although $\veff$ itself is not very illuminating, the gap equation
involving ${v_0\equiv\langle|\phi|\rangle}$,
$$
{1\over N}{\delta \veff\over \delta v_0}=0
= 2\lambda v_0-{v_0\over2\pi}
\left(\sqrt{\Lambda^2+{v_0}^2}-v_0\right),
\eqno(15)
$$
yields a non-trivial VEV for $v_0$ if
$\lambda$ is less than a critical value $\lambda_c\equiv\Lambda/{4\pi}$.
Notice that the four-Fermi coupling is $\sim 1/\lambda$ and hence
$\lambda<\lambda_c$ is a strong coupling r\'egime. This result has to be
contrasted with the BCS model where the gap arises in the weak coupling phase.
The origin of this difference lies in the behavior of the density of states.
In case of nonrelativistic fermions the density of states is taken to be
constant
near the Fermi surface whereas in the relativistic model presented here the
density of states $g(\epsilon)={\epsilon\over 4\pi}\,$goes to zero in the
$\epsilon\rightarrow 0\,$limit.

There are two
interesting limits one can consider for the solutions of the gap equation.
The first limit $\lambda\ltwid\lambda_c$
is more physical since the mean-field
approach is generally sensible near the critical point. Here,
$v_0\simeq 4\pi(\lambda_c-\lambda)$; obtaining
this solution we have assumed
that
$v_0\ll\Lambda$, and the solution indeed shows that this is self-consistent;
consequently, the mass gap in the fermion spectrum is also
small. Here, $\veff$
can also be written compactly as
$$
{\veff=N\left(\lambda-\lambda_c\right)\phi^*\phi
+{N\over{6\pi}}(\phi^*\phi)^{3/2}},
\eqno(16)
$$
from which it is evident that $\la_c$ is indeed the critical value of $\la$

The second limit corresponds to the {\sl very strong
coupling} r\'egime, $\lambda\ll\lambda_c$;
one gets from the solution of the
gap equation
$v_0\cong 2\pi\lambda_c^2/\lambda$. Here, the mass gap is large and the
fermion mass plays the role of the cut-off.

In what follows, we will confine ourselves to the first r\'egime in the
spirit of the mean-field approach.

It is worth mentioning at this point that the presence of the linear
ultraviolet divergence, although troubling at first sight,
can be handled in
large-$N$ perturbation theory,\refto{roswarpar}
where these four-Fermi couplings are
renormalizable. This cannot be done in standard perturbation theory as
seen from a simple power counting argument. In the
following section we will give
the renormalization prescription once we have the
two-loop results.

We now analyse the temperature corrections to the effective potential in
order to determine the nature of the phase transition.
In the imaginary-time formalism,\refto{matsubara} finite temperature is
achieved by making
field configurations periodic or
antiperiodic in time for bosons or fermions
respectively, with period
$\be=1/{T}$ (with the Boltzman constant put to one).
 In momentum space this amounts to replacing
$-i\int d^3k/(2\pi)^3$ by $\be^{-1}\sum_n\int d^2k/(2\pi)^2$, where
$k^0=2\pi i\be^{-1}(n+{1\over 2})$ for fermions. Finite density is
taken into account
by adding a chemical potential $\mu$ to $k^0$. The effective potential at
finite temperature and zero chemical potential is then
$$
{1\over N}\veff(\phi,\phi^*;T)=\lambda\phi^*\phi
-{1\over\be}\sum_{-\infty}^\infty\int{d^2k\over(2\pi)^2}
\log\left(1+{\phi^*\phi\over\bfk^2+{4\pi^2\over\be^2}(n+\half)^2}\right).
\eqno(17)
$$
Like the zero temperature case, we analyse the gap equation for a
non-trivial
constant solution $\vt\equiv\langle|\phi|\rangle$:
$$
\lambda \vt-{1\over \beta}\sum_{n\epsilon Z}\int{d^2k\over (2\pi)^2}
{1\over [{\bf k}^2+\vt^2+{4\pi^2\over \beta^2}(n+\half)^2]}\,=0 .
\eqno(18)
$$
After carrying out the sum over frequencies, we get
$$
\vt{\left(\lambda - \int{d^2k\over (2\pi)^2}{1\over \sqrt{{\bf k}^2+\vt^2}}
{\sinh(\beta\sqrt{{\bf k}^2+\vt^2})\over
\cosh({\beta\sqrt{{\bf k}^2+\vt^2}})+
1}\right )}=0 .
\eqno(19)
$$
Introducing the cutoff $\Lambda$ to regulate the linear divergence as for
 the
$T=0$ case
(finite temperature effects do not introduce any new ultraviolet
divergence), the
non-trivial solution $\vt\neq 0$ satisfies the equation
$$
{\lambda - {\Lambda\over 4\pi} + {1\over 4\pi\beta}
\log\left[ 2(\cosh(\beta \vt)+1)\right]
=0}.
$$
At $T=0$, we get $v_0=4\pi(\lambda_c-\lambda)$, with
$\lambda_c={\Lambda/4\pi}$, which matches with our zero-temperature
calculation.
After simplification, the finite-temperature gap equation is
$$
{1\over \beta}\log \left[ 2(\cosh(\beta \vt) + 1)\right]=v_0.
\eqno(20)
$$
The critical temperature $T_0$ is
defined to be the point where $\vt=0$, yielding
$$
T_0={v_0\over 2\log 2}.
\eqno(21)
$$
In general, allowing for a
finite
chemical potential leads to a
critical line in the $\mu - T$ plane; here the
simplified gap equation generalizes to
$$
{1\over \beta}\log \left[2(\cosh(\beta \vt)
+ \cosh(\beta \mu))\right]=v_0.
\eqno(22)
$$
The critical line in the $\mu-T$ plane is derived from
(22) by putting $\vt=0$,
and subsequently solving for $\mu$ as a function of temperature. The resulting
phase diagram is displayed in Fig.~2.

A physically relevant quantity
is the ratio of the gap parameter (at zero temperature) to the
critical temperature. From the above, we have
$2v_0/{T_0}=4\log 2\simeq2.77$, which is
less than the BCS gap-to-transition temperature
ratio of 3.52.
However, as
will be shown in Sec. IV, $T_0$ is not the superconducting transition
temperature: that temperature
will be lower than $T_0$ and
hence the gap-to-transition temperature ratio
will turn out to be larger than $2v_0/T_0$. Indeed, it will turn out to be
larger than the BCS value for $N\ltwid22$.

The neutral order parameter approaches
zero smoothly, with critical exponent
$\be=1/2$ (this is precisely the
mean-field value), so this is a second-order
phase transition.

We proceed to compute the kinetic term of the effective action.
If we are to have a sensible Ginzburg-Landau theory for the
condensate, the two-derivative term must have a positive coefficient. The
computation is very similar to that of the effective potential.
Here we present the zero-temperature case; at finite temperature, the
calculation is rendered somewhat more complicated by the fact that Lorentz
invariance is lost; this case is discussed in the Appendix.

For the kinetic term,
it is clearly Green's functions with two nonvanishing external
momenta which contribute.
In real space these terms can be arranged as in (11) to
exhibit
the charge conjugation symmetry of the original theory.
(This arrangement also makes
it easier to take into account gauge interactions by minimal substitution:
notice that the last term is already gauge covariant.)
To calculate the
constant
coefficients $Z_1$ and $Z_2$, we compute the following derivatives of the
two-derivative part of (11):
$$
\left.{\delta^2\skin \over \delta\phi(x)
\delta\phi^*(0)}\right|_{\phi=\phi^*={v_0}}=
-(Z_1+2Z_2 {v_0}^2)\partial_\mu\partial^\mu \delta^3(x) ,
\eqno(23a)
$$
and
$$
\left.{\delta^2\skin \over \delta\phi(x)
\delta\phi(0)}\right|_{\phi=\phi^*={v_0}}=
-2Z_2 {v_0}^2\partial_\mu\partial^\mu\delta^3(x)\,.
\eqno(23b)
$$
Hence,
$$
\eqalignno{
-Z_1\partial_\mu\partial^\mu\delta(x)&=
\left({\delta^2\skin \over \delta\phi(x)\delta\phi^*(0)}
-{\delta^2\skin \over \delta\phi(x)\delta\phi(0)}\right),\cr
-Z_2\partial_\mu\partial^\mu\delta(x)&={1\over 2{v_0}^2}
{\delta^2S\over \delta\phi(x)\delta\phi(0)}\,. &(24) \cr}
$$
In momentum space, the above expressions become
$$
\eqalignno{
&Z_1={1\over 6}{\partial^2\over \partial q^\mu\partial q_\mu}
\left(D^{-1}_{\phi\phi^*}(q)
-D^{-1}_{\phi\phi}(q)\right)_{q=0},\cr
&Z_2={1\over 12{v_0}^2}{\partial^2\over \partial q^\mu\partial q_\mu}
\left.\left(D^{-1}_{\phi\phi}(q)\right)\right|_{q=0}\,.&(25) \cr}
$$
Here, the $D^{-1}_{i,j}$'s are the inverse propagators for the respective
fields, which can be explicitly calculated. Consider first
$D^{-1}_{\phi\phi^*}$. It gets contributions from all
one-loop diagrams where $\phi$ and $\phi^*$ carry momenta $p$ and $-p$,
respectively; analytically, we have
$$
\eqalign{
	&{N\over 2}\tilde\phi^*(-q)\tilde\phi(q)
	\sum_n {v_0}^{2n}
	\int{d^3k\over {(2\pi)}^3}
	\sum^n_{j=0}\Tr{1\over {(\ksl+\qsl)}^{2j+1}}
	{1\over {\ksl}^{2n+1-2j}}\cr
	&=N\tilde\phi^*(-q)\tilde\phi(q)
	\int{d^3k\over (2\pi)^3}
	{k(k-q)\over {(k^2-{v_0}^2)[(k-q)^2-{v_0}^2]}}\cr
	&=i\tilde\phi^*(q)\tilde\phi(-q)D^{-1}_{\phi\phi^*}(q)\,.\cr}
$$
The above series is depicted in Fig.~3.

One finds
$$
{1\over N}D^{-1}_{\phi\phi^*}(p)= {\Lambda\over {4\pi}}
- {{v_0}\over {2\pi}} -
{-p^2+2{v_0}^2\over {4\pi \sqrt{-p^2}}}\arctan{\sqrt{-p^2}\over 2{v_0}}.
$$
Hence,
$$
\left.{\partial^2\over \partial q_\mu \partial q^\mu}
D^{-1}_{\phi\phi^*}(q)\right|_{q=0}=
{5N\over 16\pi {v_0}}
$$
A similar calculation yields
$$
\left.{\partial^2\over \partial q_\mu\partial q^\mu}
D^{-1}_{\phi\phi}(q)\right|_{q=0}=
-{N\over 16\pi {v_0}};
$$
whence
$$
\eqalignno{
&Z_1={N\over 16\pi{v_0}}\cr
&Z_2=-{N\over 192\pi{v_0}^3}\,. &(26)\cr}
$$

Incorporation of electromagnetism is trivial; since we know $\phi$ has
charge $2e$, the kinetic term must be modified to an appropriate covariant
derivative; furthermore, the last term in (11) is already gauge-covariant.
Thus, we get the zero-temperature
effective Lagrangian
$$
\LL_{\rm eff}={N\over 16\pi {v_0}}|(\partial_\mu+2i e A_\mu)\phi|^2
-{N\over 192\pi{v_0}^3}(\phi^*\partial_\mu\phi+\phi\partial_\mu\phi^*)
(\phi^*\partial^\mu\phi+\phi\partial^\mu\phi^*)
-\veff.
\eqno(27)
$$

At this point it is worth asking if a Chern-Simons
term can be present in our
effective action, thereby leading to parity and time-reversal violating
effects.
This term could arise from
one-loop diagrams with two external photons carrying momentum (in addition
to any number of
scalar fields at zero momentum). However it is easy to see that
such diagrams can never generate a Chern-Simons term. This is essentially
due to the fact that there must be an even number of external scalar legs
(to preserve gauge invariance),
and therefore an even number of fermion propagators. Since the fermion is
massless there results a trace of an even number of propagators; however
with two-component spinors a Chern-Simons
term would only arise from the trace of
three gamma matrices (proportional to $\eps_{ijk}$).
Thus, there is no induced Chern-Simons term and no spontaneous parity or
time-reversal violation, in contrast with the simplest anyon superconductivity
models\refto{wilczek4} and in accord with experiments on high-temperature
superconductors.\refto{halmarwil,keietal}

Once we have the full effective action, the expectation value of the
charged
order parameter can be computed at finite temperatures. Writing
$\phi=\rho e^{i\theta}$, it will be sufficient to  analyse the VEV of
$\phi$ when fluctuations of $\rho$ can be neglected. Hence,
$$
\langle\phi(x)\rangle=\langle\rho e^{i\theta(x)}\rangle
\simeq \vt e^{-{1 \over 2}\langle\theta^2(x)\rangle}.
\eqno(28)
$$
Going over to momentum space and expanding $\theta(\bfx,\tau)=
\sum_n\theta_n(\bfx)e^{i\om_n\tau}$, where $\om_n=2\pi n/\beta$, we have
$$
\eqalignno{
	\langle\theta^2(x)\rangle=&{1 \over \beta}\sum_{n\epsilon Z}
	\int{d^2q \over {(2\pi)}^2}\langle\theta_n({\bf q})
	\theta_{-n}{(-{\bf q})}\rangle\cr
	&={1\over \beta}\sum_n\int {d^2q\over {(2\pi)}^2}
	{D_\theta}({\bfq,n})\,.&(29)\cr}
$$
The propagator, whose derivation can be found in the Appendix, is
$$
D_\theta({\bf q},n)={8\pi \over N\vt\tanh(\beta \vt/2)}
{1\over {{\bf q}^2+\stuff({2n\pi/\beta})^2}} .
\eqno(30)
$$
Substituting this into (29), the sum can be performed, yielding
$$
\langle\theta^2\rangle=
{2\over N\vt\stuff^{1/2}
\tanh(\beta\vt/2)}\int_0^\infty dq\,{\rm coth}
{\beta q\over 2\stuff^{1/2}}.
$$
The resulting integral over $q=|\bfq|$ diverges in both the ultraviolet and
infrared; introducing respective cutoffs $\Lambda$ and $\eta$, we get
$$
\langle\theta^2\rangle=
{2\over N\vt\stuff^{1/2}
\tanh(\beta\vt/2)}
\left(\La-{2\over\beta}\stuff^{1/2}
\log\beta\eta\right),
$$
up to finite parts.

We can
eliminate the above ultraviolet divergence (a next to leading order effect)
by suitably renormalizing $\rho$,
appearing in (28).
To show
how this happens, it is sufficient to work at zero temperature
(since finite temperature effects
do not introduce any new ultraviolet divergence). The VEV of $\phi$, then,
reduces to
$$
\langle\phi\rangle=v_0e^{-{\Lambda\over Nv_0}}
$$
which up to order 1/N becomes
$$
\langle\phi\rangle\approx v_0-{\Lambda\over N}.
\eqno(31)
$$
Now recall that at the leading order $v_0=4\pi(\lambda_c-\lambda)$ with
$ \lambda_c={\Lambda\over 4\pi}$. After integrating out the $\theta$
degree of freedom, as we have done in (28), we expect the critical
coupling
to get shifted, the shift being of order 1/N in the next-to-leading order.
As will be shown explicitly in
the following section, the new critical coupling up to order 1/N
is $ \lambda^{(1)}_c=(1-{1\over N}){\Lambda\over 4\pi}$ and the new VEV is
$\langle\phi\rangle=4\pi(\lambda^{(1)}_c-\lambda)$, which precisely agrees with
the expression in (31). This justifies our assertion that the ultraviolet
divergence can be soaked up in a redifined $\rho$.

It is not possible, however, to get rid of the infrared divergence,
the dependance of $\langle\phi\rangle$ on the infrared cutoff $\eta$ is:
$$
\langle\phi\rangle={\rm const}{(\beta\eta)}^\chi,
\eqno(32)
$$
with
$\chi=2/N(\beta\vt\tanh(\beta\vt/2))$.
We see that in the limit $\eta\to 0$, $\langle\phi\rangle \to0$,
since $\chi$ is positive.
This proves that $\theta$,
the massless excitation associated with the Goldstone
mode, is responsible for the vanishing of the VEV of
the charged order parameter $\phi$ at non-zero temperatures.

\head {III. NEXT-TO-LEADING ORDER CORRECTIONS AND RENORMALIZATION}

In this section, we compute the next-to-leading
order corrections to ensure that the
quantum fluctuations do not destabilize the vacuum, either due to ultraviolet
or infrared effects. Furthermore, we
show explicitly the renormalizabilty of the theory and give a
renormalization group invariant
expression for the gap.
The approach used for this purpose is the same as the one advocated in
Ref.~\cite{roswarpar}, except that we have an additional massless field.
Interestingly, the analysis of the three point vertex shows that
it is not modified in the next-to-leading order. This is similar
to the lack of radiative corrections to the electron-phonon vertex at
low-energy, a
result known as Migdal's theorem in the literature.\refto{migdal}

For calculational convenience, we use a
reducible basis of $\gamma$-matrices, and a Majorana basis for the spinors.
A common convention for the $2\times2$ Dirac $\gamma$-matrices is
$ \gamma^0=\sigma^2, \gamma^1=i\sigma^3,
\gamma^2=i\sigma^1$, where $\sigma^i$'s are the Pauli matrices.
We can write a complex spinor in terms of two real ones as
$\psi={\chi_1+i\chi_2}.$
With $\phi=\phi_2+i\phi_1$ and $\psi^c\equiv C\bar\psi^T=\chi_1-i\chi_2$,
the fermionic part of the linearized Lagrangian in equation (9)
(restricting for simplicity to one flavour for the
moment)
can be written as:
$$
\LL=\bar\Psi i\parsl\Psi- \phi_1\bar\Psi\Psi-i\phi_2\bar\Psi
\gamma_5\Psi-\lambda N(\phi_1^2+\phi_2^2)
-e\bar\Psi\Asl \gamma_5 \Psi.
\eqno(33)
$$
Here, we have used a reducible basis involving $4\times4$
$\gamma$-matrices
$$
{\tilde \gamma}_\mu=\gamma_\mu\otimes\pmatrix{1&0\cr
					0&-1\cr},
$$
and the four-component spinor $\Psi=(\chi_1\,\chi_2)^T$.
Furthermore, the $\ga_5$ matrix has been defined as:
$$
-i\pmatrix{0&1\cr
		-1&0\cr}\,.
$$
In this formulation, the method of Refs.~\cite{aitfra,babdaspan} can
conveniently
be used to calculate the derivative expansion of the effective action.
For instance, using the fact that $\{\gamma_5,\gamma_{\mu}\}=0$,
and that $\gamma_5^2=1$, we obtain, assuming that
the $\phi^i$'s are constant,
$$
\eqalignno{
\veff=& V^0+{iN\over 2}\int{d^3p\over
(2\pi)^3}\Tr\log(\psl+\phi_1-i\phi_2\gamma_5)\cr
&=N\lambda\phi^*\phi+iN\int{d^3p\over (2\pi)^3}\log
(p^2-\phi^*\phi)\,.&(34)\cr}
$$
This expression is equivalent to the one derived in (14) modulo a $\phi$
independent term, but
appears rather easily in this approach. We have checked for consistency
that the full covariant action of the previous section follows from this
alternative method.

To show renormalizability, the bare Lagrangian can be written (neglecting
the external gauge field) with three
substraction constants in the form:
$$
\LL^{\rm (bare)}
=Z_1\bar\Psi i\parsl\Psi- Z_2\phi_1\bar\Psi\Psi-iZ_2\phi_2\bar\Psi
\gamma_5\Psi-\lambda NZ_3\left({Z_2\over Z_1}\right)^2(\phi_1^2+\phi_2^2).
\eqno(35)
$$
Here, $Z_1$, $Z_2$ and $Z_3$ are overall normalizations of the fields and
will drop out of the S-matrix.

Henceforth, we will work in Euclidean space
for convenience; the Wick-rotated functional with the fermionic sources
$\bar\eta$ and $\eta$ reads:
$$
Z(\bar\eta,\eta)=\int \DD\bar\Psi \DD\Psi \DD\phi_1\DD\phi_2
e^{-\int d^3x\,(\LL^{(E)}-\bar\eta\Psi-\bar\Psi\eta)},
\eqno(36)
$$
where
$$
\LL^{(E)}=Z_1\bar\Psi i\parsl^{(E)}\Psi+ Z_2\phi_1\bar\Psi\Psi+iZ_2\phi_2
\bar\Psi
\gamma_5\Psi+ \lambda N Z_3\left({Z_2\over Z_1}\right)^2(\phi_1^2+\phi_2^2).
\eqno(37)
$$
Without loss of generality, we can
assume that $\phi_1$ takes a VEV and hence write
$$
\eqalignno{
\phi_1=&{Z_1\over Z_2}(M+\tilde\phi_1),\cr
\phi_2=&{Z_1\over Z_2}\tilde\phi_2;&(38)\cr}
$$
where $M=M_0+{M_1/N}$ and $Z_i=1+{\hat{Z_i}/ N}$. Since we will be
computing corrections up to order 1/N, the parameters have been
conveniently expanded in the same powers . The variables
$\tilde\phi_1$ and $\tilde\phi_2$
are the fluctuations in the respective fields.

The Feynman rules can be read off easily from the Lagrangian.
Notice that to next-to-leading order we have the fermion mass insertion
$-{M_1/ N}$, given diagrammatically by $i -|- j$
and the two tadpoles
given by $-2\lambda NM_0$ and $-2\lambda\hat{Z_3}(M_0+M_1)$ arising
respectively due to the second and last term of (37).
For the proof of renormalizability, it is sufficient to show that the
full VEV
of $\phi_1$ and
that the connected Greens functions $\langle\Psi(x)\bar\Psi(y)
\rangle$, $\langle\phi_i(x)\phi_i(y)\rangle, i=1,2$,
are finite.

In the momentum space, the inverse fermionic two point
function is
$$\Gamma^{ij}(p)=Z_1\delta^{ij}\Gamma(p)\, ,$$
where
$$
\eqalignno{
\Gamma(p)&=i\psl+M_0+{M_1\over N}-{1\over N}\qquad\cr
&-{1\over N}\qquad  &(39)\cr}
$$

The dashed and the dotted lines in the above expression
correspond to the $\tilde\phi_1$ and
$\tilde\phi_2$
propagators respectively.\eject
The inverse mesonic two-point function is :
$$
\eqalignno{
\qquad &= N({Z_2\over Z_1})^2D_1^{-1}(p)+2\lambda{\hat Z_3}+2\qquad\cr
&\cr
&+2\cr
&\cr
&+2\,&(40)\cr}
$$
The vertex function is :
$$
\eqalignno{
\qquad&=-Z_2(1+{1\over N}\cr
&\cr
&+{1\over N}\qquad\qquad\qquad\qquad+\hbox{finite diagrams})&(41)\cr}$$

To leading order, the gap equation (9) can be easily derived by
demanding
that the sum of the tadpoles give zero contribution.
To next-to-leading order, the tadpole contributions diagrammatically
represented as :
\vskip 2.5cm
gives
$$
{M_1\over D_1(0)}= -2\lambda{\hat{Z_3}M_0}-{1\over 2}
\int {d^3q\over(2\pi)^3}
\left.D_{\alpha}(q;M){\partial D_{\alpha}^{-1}(q;M)\over \partial M}
\right|_{M=M_0}\, .
\eqno(42)
$$

Here,
$D_{1}^{-1}(q;M)$ and $D_2^{-1}(q;M)$ are the inverse propagators
for the $\tilde\phi_1$
and $\tilde\phi_2$ fields respectively and are given by:
$$
\eqalignno{
D_{1}^{-1}(q,M)&=2\lambda+\int {d^3p\over (2\pi)^3}
\Tr\left[{1\over (i\psl+M)}
{1\over (i\psl-i\qsl+M)}\right]\cr
&=-{M_0\over \pi}+{M\over \pi}+{(q^2+4M^2)\over (2q\pi)}
\arctan{q\over 2M}\,.&(43)\cr}
$$
and
$$
\eqalignno{
D_{2}^{-1}(q,M)&=2\lambda-\int {d^3p\over (2\pi)^3}
\Tr\left[{1\over (i\psl+M)}
\gamma_5{1\over (i\psl-i\qsl+M)}\gamma_5\right]\cr
&=-{M_0\over \pi}+{M\over \pi}+{q\over 2\pi}\arctan
{q\over 2M}\,.&(44)\cr}
$$
Note that the ultraviolet behaviour of these
propagators is given by:
$$
\left.D_{1,2}(q)\right|_{\lim q\rightarrow\infty}\sim{4\over q}\,.
\eqno(45)
$$
A straightforward calculation then yields:
$$
M_1= - 2\lambda\hat Z_3\pi-{2\Lambda\over \pi}- {4M_0\over \pi^2}\log(\Lambda)
+ \hbox{finite terms}\,.
\eqno(46)
$$
Now we consider the fermion self-energy correction due to the
$\tilde\phi_1$ exchange; using the large q behaviour of the $\tilde\phi_1$
propagator the divergent part can be identified:
$$
\eqalignno{
\qquad\equiv\Sigma_1(p)&=\int {d^3q\over (2\pi)^3}{1\over i\psl-i\qsl+M}
D_1(q)\cr
&\equiv a(0) + i\tilde a(0)\psl+ \hbox{finite terms}\, .&(47)\cr}
$$
Taking the trace yields
$$
a(0)=M_0\int {d^3q\over (2\pi)^3}{1\over q^2}{4\over q}
={2M_0\over \pi^2}\log(\Lambda)\,.
\eqno(48)
$$
Similarly, after some algebra,
$$
\eqalignno{
\tilde a(0)&=-\lim_{p\rightarrow 0}\int{d^3q\over (2\pi)^3}{1\over p^2}
{p\cdot(p-q)\over (p-q)^2+M_0^2 }\cr
&=-{2\over 3\pi^2}\log(\Lambda)\,.&(49)\cr}
$$

Writing the contribution of the $\tilde \phi_2$ field to the fermion
self-energy
as $\Sigma_2(p)=b(0)+i\tilde b(0)\psl+$ finite terms, we obtain
$b(0)=-(2M_0/ \pi^2)\log(\Lambda)$ and
$\tilde b(0)=-(2/ 3\pi^2)\log(\Lambda)$. The divergent part of the
$\tilde\phi_1$ and $\tilde\phi_2$ vertex corrections
are respectively given by $c(0)=(2/ \pi^2)\log(\Lambda)$ and
$d(0)=-(2/ \pi^2)\log(\Lambda)$.
Hence, the renormalization condition reads
$$
\eqalignno{
\hat Z_1-\tilde a(0)-\tilde b(0)&=0,\cr
\hat {Z_1}M_0+M_1-a(0)-b(0)&=0,\cr
\hat Z_2+c(0)+d(0)&=0. &(50)\cr}
$$
These conditions give
$$
\eqalignno{
\hat Z_1=&-{4\over 3\pi^2}\log(\Lambda),\cr
\hat Z_2=&0\,\cr
M_1=&{4\over 3\pi^2}M_0\log(\Lambda)\,.&(51)\cr}
$$
By using the gap equation for $M_1$, we then obtain
$$
2\lambda\hat{Z_3}=-{2\Lambda\over \pi^2} -{16M_0\over 3\pi^3}\log(\Lambda)
\eqno(52)
$$
and hence
$$
2\lambda Z_3={2\Lambda\over \pi^2}(1-{1\over N})-{M_0\over \pi} -
{16M_0\over 3\pi^3N}\log(\Lambda).
$$
At this point, with all the substraction constants properly defined, the
consistency of the renormalization procedure can be checked by computing the
inverse of the connected, mesonic two-point function and showing that it is
finite, which we now do.

The zero-momentum inverse two-point function is
$$
\eqalignno{
\quad|_{p=0}=&N\left({Z_2\over Z_1}\right)^2\left(2\lambda+
\int {d^3q\over (2\pi)^3}\Tr{1\over (i\psl+M_0)^2}\right)\cr
&+2\lambda\hat Z_3-2M_1 \int {d^3q\over (2\pi)^3}\Tr{1\over
(i\psl+M_0)^3}\cr
&+{1\over 2}\int {d^3q\over (2\pi)^3}\left[{\partial
D_{\alpha}\over \partial M}
{\partial D_{\alpha}^{-1}\over \partial M}+D_{\alpha}{\partial^2 D_{\alpha}^
{-1}\over \partial M^2}\right]_{M=M_0}\,&(53)\cr}
$$
here $\a=1,2$.

A rather lengthy calculation yields
$$
\quad|_{p=0} = {2M_0\over \pi}(\hat Z_2-\hat Z_1)+
2\lambda\hat Z_3+
{2M_1\over \pi}+{2\Lambda\over \pi^2} + \hbox{finite terms}\,.
\eqno(54)
$$
By using the
previously calculated subtraction constants $\hat Z_1$, $\hat Z_2$
and $\hat Z_3$ the divergent part of the above two-point function is zero
leaving only the finite parts.

Introducing an arbitrary subtraction point $\mu$, we get
$$
\eqalignno{
Z_1=&1 - {4\over 3\pi^2N}\log({\Lambda\over \mu})\cr
Z_2=&1\cr
2\lambda Z_3=&{2\Lambda\over \pi^2}(1 - {1\over N}) - {M_0\over \pi} -
{16M_0\over 3\pi^3N}\log({\Lambda\over \mu})\,.&(55)\cr}
$$
By demanding that the theory not depend on the scale
$\mu$, we are led
to consider $M=M(\mu)$. In particular, since $2\lambda Z_3$ should be
independent of $\mu$, we get
$$
\mu{\partial M\over \partial\mu}+{16\over 3\pi^2N}
\left(\mu{\partial M\over \partial \mu}\log{\Lambda\over \mu}
- M\right) = 0\, ,
\eqno(56)
$$
yielding $M(\mu)\sim\mu^{16/ 3\pi^2N}$.
By substituting this expression in $2\lambda Z_3$,
we get,
$$
{2\lambda Z_3\over \Lambda}={2\over \pi^2}(1-{1\over N})-
{1\over \pi}
\left({M_{\rm phys}\over \Lambda}\right)^{1-{16/ 3\pi^2N}}\,.
\eqno(57)
$$
With $\tilde \lambda\equiv{2\lambda Z_3/ \Lambda}$ and
$\tilde \lambda_c\equiv{2\over \pi^2}(1-{1/ N})$, the
$\beta$-function can be
easily computed:
$$
\beta_{\tilde \lambda}\equiv-{d\tilde \lambda\over d\log(\Lambda)}=
(1-{16\over 3\pi^2N})
(\tilde \lambda-\tilde \lambda_c)\,.
\eqno(58)
$$

\head {IV. SUPERCONDUCTIVITY}

To analyse the occurrence of superconductivity in the model
presented above,
we study the low-energy
effective action using a duality transformation.\refto{leefis,zhang}
This is necessitated by the
fact that, on a plane, the phase of a complex order parameter
is multivalued and gives rise to so-called vortex excitations; a naive
calculation fails to capture the contribution of the vortices. As will
be shown below, the description in terms of the dual
variables allows the
effective separation of the single
and the multivalued components of the complex field, and hence is
quite useful in identifying the two phases: one where vortices and
anti-vortices are tightly bound to one another,
and the other where they are free.
The phase transition between these two r\'egimes is the celebrated
Kosterlitz-Thouless transition.\refto{kostho}

In this
section we will show that in the
low-temperature r\'egime, where the vortices occur in tightly bound pairs,
the relevant contribution to the path integral comes only
from the topologically
trivial piece, leading to a ground state which exhibits superconductivity.
The photon develops a mass
in this phase and the current-current correlation
function shows a massless pole, indicative of zero resistance. When the
vortices are free,
their contribution to the path integral will be seen to
{\sl precisely negate} the contribution of the topologically trivial part,
destroying superconductivity.
This then
proves the earlier assertion that
$\tkt$ is indeed the superconducting critical
temperature.

We wish to analyse the long-wavelength London limit, neglecting fluctuations of
$|\phi|$. We therefore write $\phi=\rho e^{i\theta}$, with $\rho$ fixed at
$\vt$. As was shown in Refs.~\cite{leefis,zhang}, a direct substitution of this
change of variables into the Lagrangian (11) is inappropriate, since the
correspondence between $\phi$ and $\theta$ is not one-to-one. Rather, to
capture
the physics of vortex excitations, we make the replacement
${\partial_\mu\theta \rightarrow
\partial_\mu\theta-i\varphi^*\partial_\mu\varphi}$
with ${\varphi^*\varphi=1}$. The new variable $\theta$ is now treated as a
single-valued field and $\varphi$ describes the
vortex dynamics. One can
define a vortex current by
$$
J_\mu^{\rm vort}
= {(2\pi i)}^{-1}\epsilon_{\mu\nu\lambda}\partial^\nu{(\varphi^*
\partial^\lambda\varphi)}.
\eqno(59)
$$
This current is identically
conserved, and the corresponding
integrated charge gives the vortex number.

The dynamics is described by the finite-temperature version of the Lagrangian
(27), where the finite-temperature effective potential determines the value of
$\vt$ (see (20)), and the first term of (27) determines the Lagrangian for
$\theta,\ \phi$. (As remarked earlier, the second term is independent of these
variables.) The coefficient of the first term at finite temperature is computed
in the Appendix. For static properties we can replace $Z_1$ by its spatial
finite-temperature counterpart $Z_1^{(1)}$ given in (A.6b), so that the
Lagrangian becomes
$$
\leff={N\tanh(\beta\vt/2)\over16\pi\vt}\vt^2
\left(\partial^\mu\theta-i\varphi^*\partial^\mu\varphi+2eA^\mu\right)^2.
$$
It is useful to rewrite this in terms of an auxiliary current
$J^\mu=\partial^\mu\theta-i\varphi^*\partial^\mu\varphi+2eA^\mu$:
$$
{1\over N}\leff=-{1\over {2K}}{J_\mu}{J^\mu}
- {J_\mu}{\left(\partial^\mu\theta -
i\varphi^*\partial^\mu\varphi+2eA^\mu\right)},
\eqno(60)
$$
where $K={\vt\tanh(\beta\vt/2)/ 8\pi}$.
Since $\theta$ is now single-valued, we can integrate it out, obtaining the
constraint $\partial^\mu J_\mu = 0$;
this is readily solved by
$$
J_\mu= \epsilon_{\mu\nu\lambda}\partial^\nu a^\lambda;
\eqno(61)
$$
here $a_\mu(x)$ is a vector field which is defined globally.
In terms of this
auxiliary gauge field $a_\mu$ the new Lagrangian, after an integration by
parts, becomes:
$$
{1\over N}\leff=-{1\over {4K}}f^2_{\mu\nu}
-{2e\epsilon_{\mu\nu\lambda}a^{\mu}\partial^\nu
A^\lambda}-2\pi a^{\mu}J_{\mu}^{\rm vort},
\eqno(62)
$$
where $f_{\mu\nu}={\partial_\mu a_\nu -\partial_\nu a_\mu}$. Notice that the
above description of $J_\mu$
in terms of the gauge field $a_\mu$ brings in an unphysical
gauge degree of freedom, which must be removed by gauge fixing.

The integration over $a_\mu$ can now be performed,
and the result, in the Landau
gauge, is
$$
\leff= -2\pi^2NK J_\mu^{\rm vort}(x){g^{\mu\nu}\over {\partial^2}}
J_\nu^{\rm vort}(x)
-4\pi eNK J_{\mu}^{\rm vort}{g^{\mu\nu}\over {\partial^2}}
\epsilon_{\nu\lambda\rho}\partial^\lambda A^\rho
-e^2 NK
F^{\mu\nu}{1\over {\partial^2}} F_{\mu\nu}.
\eqno(63)
$$
The last term is the contribution of the topologically trivial piece and is
relevant for superconductivity. Explicitly,
neglecting for the moment the second term, which describes
the interaction of the
electromagnetic field with the vortex,
the current-current correlation function is:
$$
\delta^2\seff/{\delta A_{\mu}(x) \delta A_\nu(y)}={\langle j^{\mu}(x)
j^\nu(y)\rangle}
= -4e^2NK\left ({\partial^{\mu}\partial^\nu \over {\partial^2}}-
g^{\mu\nu}\right)\delta^3(x-y).
\eqno(64)
$$
This indicates a pole at zero momentum, and hence superconductivity.
To see the Meissner effect, namely the presence of a pole in the photon
propagator at nonzero monentum, we have to include the Maxwell term.
With the addition of $\LL_{\rm photon}= -{1\over 4} F_{\mu\nu}F^{\mu\nu}$,
the photon propagator does indeed have such a pole, corresponding to a
photon mass of $M_v^2={e^2Nv_0\over 2\pi}$.
Since the scalar mass near the critical point is given by
$M_s^2=8v_0^2$,
the ratio
$M_s/M_v={1\over e}\sqrt{{16\pi v_0\over N}}$ can give
information about the
nature of superconductivity (type I or type II).  However, without knowledge of
the dimensionful coupling constants $\lambda$ and $e$ (which in principle are
determined by the underlying model), we
are not able to draw any meaningful conclusion.

Now we proceed to analyse the crucial question: When
can one neglect the
vortex contribution?
As we will now show,
at low temperatures, when the vortices are confined,
their contribution to the ${F^{\mu\nu}\partial^{-2}F_{\mu\nu}}$
term is zero
and hence they do not affect superconductivity.
Physically, this is perfectly reasonable:
assuming that the interaction
of the vortices with the electromagnetic field is
small, in the confined phase a slowly varying magnetic field sees pairs of
tightly bound vortices with opposite vortex charges, and hence the net
contribution from each pair vanishes, leaving untouched the
contribution of the
single-valued part.
In contrast, above the KT
phase transition the vortex
contribution to this term is crucial:
{\sl it exactly cancels the contribution from the
single-valued part\/}, thereby
destroying superconductivity.

To see this more explicitly,
consider for simplicity only the static vortex density configurations
$(J^{\rm vort}_i=0\, \hbox{and}\, \partial_{0}J^{\rm vort}_0=0)$, and let
us compute the vortex contribution to the current-current correlation function.
Knowing that the vortex current is conserved, we can write
$$
J_0^{\rm vort}(\bfx)\equiv\rho^{\rm vort}(x)=\sum_a m_a \delta(\bfx-\bfx_a(t))
\eqno(65)
$$
where $m_a$ is the
integer-valued vorticity and $\bfx_a$ is the position of the $a^{th}$
vortex.  And making use of the green function property in two dimensions
$$
{1\over {\nabla}^2}F(\bfx)={1\over 2\pi}\int d{\bfy}\,F(\bfy) \log |\bfx-\bfy|
$$
the effective action reduces to
$$
\seff/N=\pi K \beta
\sum_{a,a'}m_a m_{a'}\log |\bfx_a-\bfx_{a'}| +
2eK \beta\sum_a m_a\int d^2r\,B(\bfx)\log |\bfx-\bfx_a|;
\eqno(66)
$$
without the interaction term this is the action of the familiar XY model.
In the static limit, the
contribution of the vortices to the current-current
correlation function is
$$
\eqalign{
	\langle j^i(\bfq) j^j(-\bfq)\rangle^{\rm vort}
        &={\delta^2 S_{\rm eff}\over
	\delta A^i(\bfq)\delta A^j(-\bfq)}\cr
	&=\left({\delta^{ij}-
	{q^iq^j\over \bfq^2}}\right)\bfq^2
        \langle\rho^{\rm vort}(\bfq)\rho^{\rm vort}(-\bfq)\rangle.\cr}
\eqno(67)
$$
In real space,
$$
 \rho^{\rm vort} (\bfq)=\sum_{a}m_{a}e^{i\bfq\cdot \bfx_{a}}
$$
and hence in the confined phase
$$
\langle \rho^{\rm vort}(\bfq)\rho^{\rm vort}(-\bfq)\rangle
=\sum_{a,a'}\langle e^{i\bfq\cdot(\bfx_a-\bfx_{a'})}
\rangle=-2\langle e^{i\bfq\cdot \bfR}\rangle + (m_a=2,3\dots).
\eqno(68)
$$

The expectation value of $\langle e^{i\bfq\cdot \bfR}\rangle$
can be easily computed using
the action for the XY model,\refto{joskadkirnel} yielding
$$
\langle e^{i\bfq\cdot \bfR}\rangle \sim {1\over\bfq^{2-2P}}
$$
and
$$
\langle \rho^{\rm vort}(\bfq)\rho^{\rm vort}(-\bfq)\rangle
\sim  {1\over \bfq^{2-2P}};
\eqno(69)
$$
here $P=\pi NK\beta$.
In the confined phase $P>2$, $\langle \rho^{\rm vort}(\bfq)
\rho^{\rm vort}(-\bfq)\rangle\sim \bfq^{p-2}$,
where $p>4$. With (67), this implies that the zero
momentum pole in $\langle j^i(\bfq)j^j(-\bfq)\rangle$ does not get a
contribution from the confined vortices. This then implies that in
$\leff$ the $ F^{\mu\nu}\partial^{-2}F_{\mu\nu}$ term and hence the
Meissner effect is not affected by the vortex sector.

As an aside, it is perhaps worth pointing out that in position space
$$
\langle \rho^{\rm vort}(\bfx)\rho^{\rm vort}(\bfy)\rangle\sim |x-y|^{-2P},
$$
which is the
well-known power law behaviour of the correlation functions in the confined
phase.

In contrast, at temperatures
above the vortex confinement-deconfinement transition,
when the vortices are free and the interaction
between them is screened as in a plasma, we expect a rather substantial
contribution from that sector. In this phase directly integrating out the
$J_\mu$ variable yields:
$$
\leff^{\rm vort}= NKe^2F^{\mu\nu}{1\over \partial^2}F_{\mu\nu}
\eqno(70)
$$
which is {\sl precisely of opposite sign} to that of the previously computed
contribution of the single-valued sector (\cf (63)).
This substantiates the physical
argument given above that in the deconfined phase superconductivity
is destroyed and $T_c=\tkt $. This is one of the main results of this
paper.

We note that the same mechanism has been invoked in other models
\refto{kovros,dormav} to explain
superconductivity.  In these works the logarithmic confining potential was
due to the exchange of massless gauge field, however the disappearance of the
Meissner effect above $T_{KT}$ has not been shown explicitly.
Our analysis may also be applicable to these models; this is currently under
investigation.

The confining-deconfining phase transition in this model occurs at
$\pi K N\beta=2$, or,
$$
{\beta\vt\over 2}\tanh{\beta\vt\over 2}={8\over N}.
\eqno(71)
$$
Here $\vt$, the expectation value of $|\phi|$ at finite temperature, is given
by
(20). Eqs.~(20) and (71)
can (in principle) be solved for the KT transition
temperature $\tkt $, which as explained above is also the superconducting
transition temperature. It is not difficult to see that for any $N$ there is a
unique solution to (71); indeed, we display the left side of (71) as a function
of inverse temperature in Fig.~4. Its intersection with the line $8/N$ gives
the
KT transition temperature. Clearly $\tkt $
increases with $N$; its minimum value
is obviously $\beta_0$, the inverse of the critical temperature for the
transition wherein $|\phi|$ obtains an expectation value.

We are now in a position to estimate the
ratio of the gap parameter to the superconducting critical temperature,
$2v_0/T_c=2v_0/\tkt $, as a function of $N$.
This ratio is a decreasing function of $N$,
with minimum value $4\log2\simeq2.77$ as $N\to\infty$. For $N$ small (more
precisely, for values of $N$ for which $8/N\gg1$), the ratio tends to the
asymptotic form $32/N$. (Since $N$ is integral, this asymptotic form is
reasonable for $N\leq3$.)
In many physically interesting
models\refto{maraff} related
to the Hubbard model, $N=2$, in which case the above ratio is approximately
16,
considerably larger than the BCS value of 3.52.
In fact, it is found numerically that the ratio is larger than the BCS value
for $N\ltwid22$.

\head {V. CONCLUSIONS}

To conclude, we have presented a relativistic model in $2+1$ dimensions
which exhibits superconductivity up to a critical temperature
$T_c$; unlike the anyonic theories, the ground state does not violate
parity or time-reversal symmetries. Interestingly, the gap-to-$T_c$ ratio
is higher than the conventional BCS value. It should be emphasized that
relativistic field theories are not uncommon in condensed matter systems;
quasi-particles satisfying linear dispersion relations are described by such
theories. Prominent examples appear in magnetic systems, $\it e.g$, the
$0(3)$
sigma model descibes the low-energy modes of the antiferromagnetic spin
system and Dirac fermions are the relevant excitations of the so called
flux-phase ground state. These models are widely believed to be relevant for
high-$T_c$ materials because of their connection with the Hubbard model.

The KT mechanism of vortex binding played a crucial role in the
occurrence of superconductivity
in our model. These vortices arise due to the
multivaluedness of the phase degree of freedom of the order parameter. We
have explicitly demonstrated
how the photon mass vanishes once the vortices unbind above some critical
temperature. Concurrently, the pole in the current-current correlator
disappears. We note that this analysis can also be applied to other
models\refto{kovros,dormav,leefis,senechal} and
to effective field theories describing the quantized Hall
effect.\refto{zhang}
Furthermore,
we remark that the relevance of the KT transitions to quasi-two-dimensional
systems has been
discussed previously;\refto{hiktsu} such an extension to the model
described here can also be carried out. Work along
these lines is currently under progress and will be reported elsewhere.

Another crucial ingredient of our
analysis is the large-$N$ perturbation theory;
this technique has been extensively used in condensed matter physics.
In fact
it has been
recently applied to the $0(3)$ sigma model leading to interesting
results.\refto{chahalnel}
It should be noted that interesting gap-to-$T_c$ ratio is reached for
$N\ltwid22$ in our model; it is amusing that multiplets of fermions appear in
the low-energy dynamics of Hubbard and related models.\refto{maraff}
Hence, the precise
connection of the model elucidated here to that of the
microscopic lattice Hamiltonians needs further scrutiny to establish the
values of the parameters used and the relevant
low-energy degrees of freedom.

We acknowledge useful discussions with
V. Spiridonov, D. Arovas, B. Sakita,
M.B. Paranjape and R.
Ray, and technical help from Dr.P.
Tripathy in the drawings of diagrams.
We also thank K. Shizuya for many thoughtful comments on an earliar
version of this manuscript. R.M. thanks the Theory Group of
Lawrence Berkeley Laboratory for their kind hospitality while this work was
being completed.
This work was supported in part by the
Natural science and
Engineering Research Council of Canada and the Fonds F.C.A.R. du
Qu\'ebec.
\vfill\eject
\head {APPENDIX}

Here, we present the finite temperature propagator for
$\theta(x)$ when $|\phi|$ takes a
non-vanishing VEV. Since Lorentz invariance is lost when $T\neq0$, the two
derivative part of the effective action has to be separated in to
temporal and spatial parts; the coefficients are different at non-zero
temperatures. The two derivative part can then be written as:
$$
\LL^{\rm kin}
=Z_1^{(0)}\partial_0\phi\partial_0\phi^* -
Z_1^{(1)}\nabla\phi\cdot\nabla\phi^*+Z_2^{(0)}(\phi^*\partial_0\phi+
\phi\partial_0\phi^*)^2-Z_2^{(1)}(\phi^*\nabla\phi+\phi\nabla
\phi^*)^2\,.
\eqno(A.1)
$$
For low-energy calculations we neglect the
fluctuations in the magnitude of $\phi$, writing
$\phi=(\vt) e^{i\theta}$. Then the last two terms drop out and the kinetic term
becomes
$$
\LL^{\rm kin}=\vt^2\left[Z_1^{(0)}(\partial_0\theta)^2-Z_1^{(1)}
(\nabla\theta)^2\right]\,.
\eqno(A.2)
$$
After performing an integration by parts:
$$
S(\theta)=\int d^3x\,{1\over 2}\theta(x) D^{-1}_{\theta}\theta(x)\,.
\eqno(A.3)
$$
Here, the inverse propagator for $\theta(x)$ is given by:
$$
D^{-1}_{\theta}(\bfq;\omega)=2\vt^2
\left(Z_1^{(0)}\omega^2+Z_1^{(1)}\bfq^2\right)\,.
\eqno(A.4)
$$
To calculate the $Z_1^{(i)}$'s and $Z_2^{(i)}$'s we proceed as for the zero
temperature case. We notice that
$$
{\delta^2 S\over \delta\phi(x)\delta\phi^*(0)}
=\left(-Z_1^{(0)}\partial_0^2+Z_1^{(1)}\nabla^2
-2Z_2^{(0)}\vt^2\partial^2_0
+2Z_2^{(1)}\vt^2\nabla^2\right)\delta^3(x)\,.
\eqno(A.5a)
$$
Similarly,
$$
{\delta^2 S\over \delta\phi(x)\delta\phi(0)}=\left(-2Z_2^{(0)}{\vt}^2
\partial^2_0+2Z_2^{(1)}{\vt}^2\nabla^2\right)\delta^3(x)\,.
\eqno(A.5b)
$$
Although we are in fact only interested in $Z_1^{(i)}$, they are not isolated
in
(A.5a) and we must calculate $Z_2^{(i)}$ as well.

A straightforward calculation gives
$$
\eqalignno{
Z_1^{(0)}&=-N{\beta \vt-\sinh(\beta \vt)\over
32\pi \vt\cosh^2({\beta \vt/2})}\cr
&={N\over 16\pi \vt}\left(\tanh({\beta \vt/2})-
{\beta\vt\over 2\cosh^2({\beta \vt/2})}\right)\,&(A.6a)\cr}
$$
and
$$
Z_1^{(1)}={N\over 16\pi \vt}\tanh({\beta \vt/2})\,.
\eqno(A.6b)
$$
Similarly,
$$
Z_2^{(0)}={N\over 192\pi\vt^3}\left({(\beta\vt)^2\tanh({\beta \vt/2})
\over \cosh^2({\beta \vt/2})}
+\left({\beta \vt\over 2}{1\over \cosh^2({\beta \vt/2})}
-\tanh({\beta \vt/2})\right)\right)\,
\eqno(A.7a)
$$
and
$$
Z_2^{(1)}={N\over 192\pi \vt^3}\left[{\beta \vt\over 2\cosh^2({\beta \vt/2})}
-{\tanh({\beta \vt/2})}\right]\,.
\eqno(A.7b)
$$
Using these expressions, we get
$$
D_{\theta}(\bfq^2;\omega)={8\pi \over N\vt\tanh({\beta \vt/2})}{1\over
\stuff \omega^2+\bfq^2}\,.
\eqno(A.8)
$$
Identifying the frequency $\omega$ with the Matsubara frequency $\omega_n=2\pi
n/\beta$, we obtain (30), as desired.

\figurecaptions

1. One-loop contribution to $\veff$.

2. Critical line in the $\mu-T$ plane.

3. One-loop diagrams contributing to the two-derivative term in $\seff$.

4. Graphical solution of Eq.~(71).

\vfill
\endfigurecaptions
\eject
\references


\refis{affzouhsuand}
I. Affleck, Z. Zou, T. Hsu and P.W. Anderson,
\prb 38, 745, 1988.

\refis{ahaboh}
Y. Aharonov and D. Bohm, \pr 115, 485, 1959.

\refis{aitfra}
I.J.R. Aitchison and C.M. Fraser, \pl 146B, 63, 1984.

\refis{alfwil}
M. Alford and F. Wilczek, \prl 62, 1071, 1989.

\refis{anderson}
P.W. Anderson, \journal{Science}, 235, 1196, 1987.

\refis{aroschwilzee}
D.P. Arovas, \etalc \np B251 [FS13], 117, 1985.


\refis{babpanram} K.S. Babu, P.K. Panigrahi and S. Ramaswamy,
\prd 39, 1190, 1989.

\refis{babdaspan} K.S. Babu, A. Das and P.K. Panigrahi,
\prd 36, 3725, 1987.

\refis{balercmorsri}
A.P. Balachandran, E. Ercolessi, G. Morandi and A.M. Srivastava,
{\sl Hubbard Model and Anyon Superconductivity} (World Scientific, 1990).

\refis{banlyk}
T. Banks and J. Lykken, \np B336, 500, 1990.

\refis{barcoosch}
J.~Bardeen, L.N. Cooper and J.R. Schrieffer, \pr 108, 1175, 1957.

\refis{barchodreweiyan}
W. Bardeen, \etalc \prd 11, 1094, 1975.

\refis{barelifrirab}
W.A.~Bardeen, \etalc \np B218, 445, 1983.

\refis{berezinski}
V.L. Berezinski, \journal{Sov. Phys. JETP}, 32, 493, 1970.

\refis{blaboy}
R. Blankenbecler and D. Boyanovsky, \prd 34, 612, 1986.

\refis{bowkarwij}
M.J. Bowick, D. Karabali and L.C.R. Wijewardhana,
\np B271, 417, 1986.

\refis{boyblayah}
D. Boyanovsky, R. Blankenbecler and R. Yahalom,
\np B270 [FS16], 483, 1986.

\refis{boyjas}
D. Boyanovsky and D. Jasnow, \journal Physica, A177, 537, 1991.

\refis{bradavmat}
R. H. Brandenberger, A.-C. Davis and A. M. Matheson,
\np B307, 909, 1988.

\refis{brapri}
R.A. Brandt and J.R. Primack,
\journal{Int.~J.~Theor.~Phys.}, 17, 267, 1978.

\refis{brown}
L.S. Brown, \prl 57, 499, 1986.


\refis{callan1}
C.G. Callan, Jr, \prd 25, 2141, 1982.

\refis{callan2}
C.G.~Callan, Jr, \pr D26, 2058, 1982.

\refis{callan3}
C.G. Callan, Jr, \np B212, 391, 1983.

\refis{calhar}
C.G. Callan, Jr and J. Harvey, \np B250, 427, 1985.

\refis{campbell1}
D.K.~Campbell, \pl 64B, 187, 1976.

\refis{campbell2}
D.K.~Campbell, \annp 129, 249, 1980.

\refis{campbell3}
D.K.~Campbell, \prl 50, 865, 1983.

\refis{cambis}
D.K.~Campbell and A.R.~Bishop, \np B200 [FS4], 297, 1982.

\refis{cambisric}
D.K.~Campbell, A.R.~Bishop and T.~M.~Rice, in {\sl
Handbook of Conducting Polymers, vol 2}, ed. T.A.~Skotheim, Dekker, 1986.

\refis{camlia}
D.K.~Campbell and Y.-T.~Liao, \prd 14, 2093, 1976.

\refis{cardy1}
J.L. Cardy, \np B205 [FS5], 17, 1982.

\refis{carvelwagcha}
M. Carena, W.T.A. ter Veldhuis, C.E.M. Wagner and S.
Chaudhuri, \pl B252, 101, 1990.

\refis{chahalnel}
S. Chakravarty, B. Halperin and D. Nelson,\prb 39, 2344, 1989.

\refis{chewilwithal}
Y-H. Chen, F. Wilczek,
E. Witten and B. Halperin, \ijmp B3, 1001, 1989.

\refis{chewil}
Y-H.Chen and F.Wilczek, \ijmp B3, 117, 1989.

\refis{cjjtw}
A. Chodos, \etalc \prd 9, 3471, 1974.

\refis{coleman1}
S.~Coleman, \prd 11, 2088, 1975.

\refis{coleman2}
S.~Coleman, \np B262, 263, 1985.

\refis{coleman3}
S.~Coleman, \journal {Commun. Math. Phys.}, 31, 259, 1973.

\refis{coleman4}
S.~Coleman,
{\sl Aspects of Symmetry} (Cambridge University Press, 1985).

\refis{coljacsus}
S.~Coleman, R.~Jackiw and L.~Suskind,
\annp 93, 267, 1975.

\refis{colwei}
S.~Coleman and E.~Weinberg, \prd 7, 1888, 1973.

\refis{copkollee}
E.~Copeland, E.W.~Kolb and K.~Lee, \pr D38, 3023, 1988.

\refis{creutz}
M. Creutz, \prd 10, 1749, 1974.

\refis{cresoh}
M. Creutz and K.S. Soh, \prd 12, 443, 1975.


\refis{dashasnev1}
R.F.~Dashen, B.~Hasslacher and A.~Neveu,
\prd 10, 4114, 1974.

\refis{dashasnev2}
R.F.~Dashen, B.~Hasslacher and A.~Neveu,
\prd 10, 4130, 1974.

\refis{dashasnev3}
R.F.~Dashen, B.~Hasslacher and A.~Neveu,
\prd 10, 4136, 1974.

\refis{dashasnev5}
R.F.~Dashen, B.~Hasslacher and A.~Neveu,
\prd 12, 2443, 1974.

\refis{desjactem}
S. Deser, R. Jackiw, S. Templeton, \annp 140, 372, 1982.

\refis{doljac}
L.Dolan and R.Jackiw, \prd 9, xxxxxxxxxxxxx, 1974.***Check***

\refis{dormav}
N. Dorey and N.E. Mavromatos, \np B386, 614, 1992.

\refis{domrea}
T. Dombre and N. Read, {\sl On the
absence of the Hopf invariant in the long wavelength action of
2D quantum antiferromagnets}, MIT preprint (1988).

\refis{dzhpolwei}
I. Dzhaloshinskii, A.M. Polyakov and P. Wiegmann,
\pl 127A, 112, 1988.



\refis{fethanlau}
A.L. Fetter,
C.B. Hanna and R.B. Laughlin, \prb 39, 9679, 1989.

\refis{feynman}
R. P. Feynman, {\sl Statistical Mechanics} (W. A.
Benjamin, Reading, MA, 1972).

\refis{feyleisan}
R.P. Feynman, R.B. Leighton and M. Sands,
{\sl The Feynman Lectures on Physics}, vol. 3 (Addison-Wesley, 1963-65).

\refis{fradkin}
E. Fradkin, {\sl Field Theories of Condensed Matter Systems} (Addison-Wesley,
1991).

\refis{frasto} 
E. Fradkin and M. Stone,
{\sl Topological terms in one- and two-dimensional quantum
Heisenberg antiferomagnets}, Illinois preprint P/88/4/44.

\refis{frilee1}
R.~Friedberg and T.D.~Lee, \prd 15, 1694, 1977.

\refis{frilee2}
R.~Friedberg and T.D.~Lee, \prd 16, 1096, 1977.

\refis{frileesir1}
R.~Friedberg, T.D.~Lee and A.~Sirlin, \prd 13, 2739, 1976.

\refis{frileesir2}
R.~Friedberg, T.D.~Lee and A.~Sirlin,
\np B115, {1, 32}, 1976.


\refis{ganhat}
N. Ganoulis and M. Hatzis, \pl 198B, 503, 1987.

\refis{gatkovros}
G. Gat, A. Kovner and B. Rosenstein,
\np B385, 76, 1992.

\refis{ganpan}
A. Gangopadhyaya and P.K. Panigrahi, \prb 44, 9749, 1991.

\refis{gersin}
C. C. Gerry and V. A. Singh, \prd 20, 2250, 1979.

\refis{gineps}
J.M.~Ginder and A.J.~Epstein, \prb 41, 10674, 1990.

\refis{goldhaber1}
A.S. Goldhaber, \prl 36, 1122, 1976.

\refis{goldhaber2}
A.S. Goldhaber, \prl 49, 905, 1982.

\refis{golmac}
A.S. Goldhaber and R. MacKenzie, \pl 214B, 471, 1988.

\refis{golmacwil}
A.S. Goldhaber, R. MacKenzie, F. Wilczek,
\mpl A4, 21, 1989.

\refis{golmensha}
G.A. Goldin, R. Menikoff and D.H. Sharp,
\journal{J.~Math.~Phys.}, 22, 1664, 1981.

\refis{gronev}
D.J. Gross and A. Neveu, \prd 10, 3235, 1974.


\refis{hagen1}
C.R. Hagen, \journal{Ann. Phys.}, 157, 342, 1984.

\refis{hagen2}
C.R. Hagen, \prd 41, 2015, 1990.

\refis{haldane}
F.D.M. Haldane, \prl 61, 1029, 1988.

\refis{halmarwil}
B.I. Halperin, J. March-Russel and F. Wilczek, \prb 40, 8726, 1989.

\refis{hanroczahzha}
T.H. Hansson, \etalc \pl 214B, 475, 1988.

\refis{hiktsu}
S. Hikami and T. Suneto, \journal{Prog. Theor. Phys.}, 63, 387, 1980.

\refis{hohenberg}
P.C. Hohenberg, \pr 153, 493, 1967.

\refis{holdom}
B.~Holdom, \pr D36, 1000, 1987.

\refis{hosotani}
Y. Hosotani, UMN-TH-1106 (1992).

\refis{hoscha}
Y. Hosotani and S. Chakravarty, \prb 42, 342, 1990.

\refis{hubbard}
J.~Hubbard, \prb 17, 494, 1978.

\refis{hubbard1}
J. Hubbard, \prl 3, 77, 1959.

\refis{hugpol}
J. Hughes and J. Polchinski, \np B278, 147, 1986;



\refis{jacreb1}
R.~Jackiw and C.~Rebbi, \prd 13, 3398, 1976.

\refis{jacred}
R. Jackiw and A.N. Redlich, \prl 50, 555, 1983.

\refis{jacros}
R. Jackiw and P. Rossi, \np B190 [FS3], 681, 1981.

\refis{jacsch}
R.~Jackiw and J.~R.~Schrieffer, \np B190 [FS3], 253, 1981.

\refis{joskadkirnel}
J.V. Jos\'e, L.P. Kadanoff, S. Kirkpatrick and D.R. Nelson,
\prb 16, 1217, 1977.


\refis{kallau}
V. Kalmeyer and R.B. Laughlin, \prl 59, 2095, 1987.

\refis{kostho}
J.M. Kosterlitz and D.J. Thouless, \journal {J. Phys. C}, 6, 1181, 1973.

\refis{kovros}
A. Kovner and B. Rosenstein, \prb 42, 4748, 1990.

\refis{kieetal}
R.F. Kiefl, \etalc \prl 64, 2082, 1990.

\refis{kriroz}
I.V.~Krive and A.S.~Rozhavski{\u\i},
\journal Sov. Phys. Usp., 30 (5), 370, 1987.

\refis{krusch}
J.A.~Krumhansl and J.R.~Schrieffer, \prb 11, 3535, 1975.


\refis{laimor}
M.~G.~G.~Laidlaw and C.~Morette DeWitt, \prd 3, 1375, 1971.

\refis{laughlin}
R.B. Laughlin, \prl 60, 2677, 1988.

\refis{laughlin1}
R.B. Laughlin, \journal {Science}, 242, 525, 1988.

\refis{leefis}
D.-H. Lee and M.P.A. Fisher, \prl 63, 903, 1989.

\refis{leepan}
T.D.~Lee and Y.~Pang, \np B315, 477, 1989.

\refis{leewic}
T.D. Lee and G.C. Wick, \prd 9, 2291, 1975.

\refis{leinaas}
J.M. Leinaas, \journal{Nuo. Cim.}, A47, 19, 1978.

\refis{leimyr}
J.M. Leinaas and J. Myrheim,
\journal{Nuo. Cim.}, B37, 1, 1977.


\refis{mackenzie2}
R. MacKenzie, \np B303, 149, 1988.

\refis{macpansak}
R. MacKenzie, P.K. Panigrahi and S. Sakhi,
to appear, Phys. Rev. B.

\refis{maceps}
A.G.~MacDiarmid and A.J.~Epstein,
\journal{Farad. Discuss. Chem. Soc.}, 88, 317, 1989.

\refis{macmat}
R. MacKenzie and A. Matheson, \pl 259B, 63, 1991.

\refis{macwil6}
R. MacKenzie and F. Wilczek, \ijmp A3, 2827, 1988.

\refis{macwilzee}
R.~MacKenzie, F.~Wilczek and A.~Zee, \prl 53, 2204, 1984.

\refis{mandelstam}
S.~Mandelstam, \prd 11, 3026, 1975.

\refis{maraff}
J.B.Marston and I.Affleck, \prb 39, 11538, 1989.

\refis{marwil}
J. March-Russell and F. Wilczek, \prl 61, 2066, 1988.

\refis{matsubara}
T. Matsubara, \journal {Prog.Theor. Phys.}, 14, 351, 1955.

\refis{mccmac}
J. McCabe and R. MacKenzie, Montreal preprint
UdeM-LPN-TH-102.

\refis{merwag}
N.D. Mermin and H. Wagner, \prl 17, 1133, 1966.

\refis{migdal}
A.B. Migdal, \journal{Sov. Phys. JETP}, 7, 996, 1958.

\refis{mori}
H. Mori, \prb 42, 184, 1990.

\refis{moran}
S.~Moran, {\sl The Mathematical Theory of Knots and Braids:
An Introduction} (North Holland, Amsterdam, 1983).


\refis{namjon}
Y. Nambu and G. Jona-Lasinio, \pr 122, 345, 1961;
\journal{}, 124, 246, 1961.

\refis{nieole}
H.B. Nielsen and P. Olesen, \np B61, 45, 1973.

\refis{niesem1}
A. Niemi and G.W. Semenoff, \prl 51, 277, 1983.



\refis{panraysak}
P.K. Panigrahi, R. Ray and B. Sakita, \prb 42, 4036, 1990.

\refis{paranjape1}
M.B. Paranjape, \prl 55, 2390, 1985
(erratum: \prl 57, 500, 1986).

\refis{paranjape2}
\prd 36, 3766, 1987.

\refis{paukha}
S.K. Paul and A. Khare, \pl 174B, 420, 1986.

\refis{perry}
R.J.~Perry, PhD thesis, University of Maryland, 1984.

\refis{peshkin}
\journal {Phys.~Rep.}, 80, 375, 1981.

\refis{polaron}
Many aspects of the polaron problem in solid state physics,
as well as references to the original literature, can be found in {\sl
Polarons and excitons}, Proceedings, 1962 Scottish Universities' Summer
School in Physics,  eds. C.G.~Kuper and G.D.~Whitfield, Plenum Press, 1963.

\refis{polyakov1}
A.M. Polyakov, \mpl A3, 325, 1988.

\refis{polychronakos}
A.P. Polychronakos, \np B278, 207, 1986.

\refis{preskill}
J.~Preskill, ``Vortices and Monopoles,'' in {\sl Architecture of fundamental
interactions at short distances, Les Houches, session XLIV, 1985\/}, P. Ramond
and R. Stora, eds. (Elsevier Science Publishers, 1987).



\refis{rajaraman}
R.~Rajaraman, {\sl Solitons and Instantons},
North-Holland, 1982.

\refis{ransalstr}
S. Randjbar-Daemi, A. Salam and J. Strathdee, \np B340, 403, 1990.

\refis{ricmel}
M.J.~Rice and E.J.~Mele, \prb 25, 1339, 1982.

\refis{roswarpar}
B. Rosenstein, B.J. Warr and S.H. Park, \journal Phys.
Rep., 205, 59, 1991.

\refis{rubakov1}
V. Rubakov, \journal{Pis'ma Zh. Eksp. Theor. Fis.~},
33, 658, 1981 [\journal{JETP Lett.}, 33, 644, 1981].

\refis{rubakov2}
V. Rubakov, \np B203, 311, 1982.


\refis{sakita}
B. Sakita, {\sl Quantum theory of many-variable systems and
fields} (World Scientific, 1985).

\refis{schakel}
A.M.J. Schakel, {\sl On broken symmetries in Fermi systems},
University of  Amsterdam thesis (1989).

\refis{schakel1}
A.M.J. Schakel, \prd 44, 1198, 1991.

\refis{schonfeld}
J.F. Schonfeld, \np B185, 157, 1981.

\refis{schul1}
L.~S.~Schulman, \pr 176, 1558, 1968.

\refis{schul2}
L. S. Schulman, \pr 188, 1139, 1969.

\refis{schul3}
L.S. Schulman, {\sl Techniques and Applications of
Path Integration} (Wiley, New York, 1981).

\refis{semwei}
G.W. Semenoff and N. Weiss, \pl 250B, 117, 1990.

\refis{semwij}
G.W. Semenoff and L.C.R. Wijewardhana, \pl 184B, 397, 1987.

\refis{semwij1}
G.W. Semenoff and L.C.R. Wijewardhana, \prl 63, 2633, 1989.

\refis{senechal}
D. S\'en\'echal, \pl 297B, 138, 1992.

\refis{shawil}
A. Shapere and F. Wilczek, \np B320, 669, 1989.

\refis{shankar}
R.~Shankar, \np B330, 433, 1990.

\refis{shawit}
R.~Shankar and E.~Witten, \np B141, 349, 1978.

\refis{siegel}
W. Siegel, \np B156, 135, 1979.

\refis{staforaya}
P.C.E. Stamp, L. Forro and C. Ayache, \prb 38, 2847, 1988;
S. Martin \etalc \prl 62, 677, 1989.

\refis{stratonovich}
R.L. Stratonovich, \journal Dokl. Akad. Nauk. SSSR, 115, 1097, 1957.

\refis{suschhee1}
W.P.~Su, J.R.~Schrieffer, and A.J.~Heeger,
\prl 42, 1698, 1979.

\refis{suschhee2}
W.P.~Su, J.R.~Schrieffer, and A.J.~Heeger,
\prb 22, 2099 1980 (erratum: \prb 28, 1138, 1983).

\refis{tsukan}
A. Tsuchiya and Y. Kanie,
\journal Lett.~Math.~Phys., 13, 303, 1987.

\refis{suzsribhalaw}
A. Suzuki, M.K. Srivastava, R.K. Bhaduri and J. Law,
\prb 44, 10731, 1991.


\refis{thooft}
G. 't Hooft, \np B72, 461, 1974.

\refis{trugenberger}
C. Trugenberger, \prd 45, 3807, 1992.

\refis{trugenberger1}
C. Trugenberger, \pl 288B, 121, 1992.



\refis{vilenkin}
A. Vilenkin, \journal Phys. Rep., 121, 263, 1985.

\refis{vinciarelli}
P. Vinciarelli,
\journal Nuo. Cim. Lett., 4, 905, 1972.


\refis{weinberg}
E.J. Weinberg, \prd 24, 2669, 1981.

\refis{weiss}
N. Weiss, \mpl A7, 2627, 1992.

\refis{wenwilzee}
X.G. Wen, F. Wilczek and A. Zee, \prb 39, 11413, 1989.

\refis{wenzee}
X.-G. Wen and A. Zee, \prl 61, 1025, 1988.

\refis{wenzee1}
X.G.Wen and A.Zee, \prb 44, 274, 1991.

\refis{wenzee2}
X.G.Wen and A.Zee, \prb 41, 240, 1990.

\refis{wiegmann}
P. B. Wiegmann, \prl 60, 821, 1988.

\refis{wilczek1}
F. Wilczek, \prl 48, 1144, 1982.

\refis{wilczek2}
F. Wilczek, \prl 48, 1146, 1982.

\refis{wilczek3}
F. Wilczek, \prl 49, 957, 1982.

\refis{wilczek4}For references and review, see,
F. Wilczek, {\sl Fractional statistics and anyon superconductivity}
(World Scientific, 1990).

\refis{wilzee1}
F. Wilczek and A. Zee, \prl 51, 2250, 1983.

\refis{witten1}
E. Witten, \pl 86B, 283, 1979.

\refis{witten2}
E. Witten, \np B249, 557, 1985.

\refis{wu1}
Y.-S. Wu, \prl 52, 2103, 1984.

\refis{wu2}
Y.-S.~Wu, \prl 53, 111, 1984.



\refis{yamich}
H. Yamamoto and I. Ichinose, \np B370, 695, 1992.

\refis{zee}
A. Zee, ``Semionics: A Theory of High Temperature
Superconductivity'', in {\sl High Temperature Superconductivity\/}, Eds.
K.~Bedell, D.~Pines and J.R.~Schrieffer (Addison-Wesley, 1990).

\refis{zhang}
S.C. Zhang, \ijmp B6, 25, 1992.

\refis{zou}
Z.Zou, \prb 40, 2262, 1989.

\endreferences

\bye